\documentclass[aps,preprint,showpacs,preprintnumbers,amsmath,amssymb]{revtex4}
\usepackage[dvips]{graphicx}
\usepackage{float}
\begin{document}

\title{Production of Z bosons and neutrinos in early universe}
\author{Cosmin Crucean \thanks{E-mail:~~crucean@physics.uvt.ro}
\\{\small \it Faculty of Physics, West University of Timi\c soara,}\\
{\small \it V. Parvan Avenue 4 RO-300223 Timi\c soara,  Romania}}

\begin{abstract}
Production of Z bosons and neutrinos is studied in the expanding de Sitter universe. The expression of the transition amplitudes in the case of Z boson interaction with leptons is established by using perturbative methods. Then the amplitude and probability for the spontaneous generation from vacuum of a Z boson a neutrino and an antineutrino are computed analytically and a graphical analysis is performed in terms of the expansion parameter. We found that the probability for this process is nonvanishing only for large expansion conditions of the early Universe. We discuss the Minkowski limit and obtain that in this limit the amplitude is zero, result which corresponds to the well established fact that spontaneous particle generation from vacuum in Minkowski space-time is forbidden by the simultaneous energy and momentum conservation in perturbative processes. The total probability of the process is computed and we prove that this quantity is important only for the regime of large expansion from early universe and is vanishing in the Minkowski limit.
\end{abstract}

\maketitle

\section{Introduction}
One of the well established theories from physics is the electro-weak theory which combines interactions between massive bosons and Dirac fermions \cite{3,4,5,6,7,8,9,10,11,cr,12}. Since it is known that the massive bosons were produced in the early universe, it is a matter of profound importance to understand the mechanism that generate these bosons. The present knowledge is that if we look earlier and earlier in times, it is
known that the temperatures and densities were so high that these bosons could be present in large numbers at thermal equilibrium \cite{w1,w2}. These bosons could also interact with other particles. Taking into account the strong gravitational fields of the early universe, then the massive bosons could be produced in perturbative processes, giving rise to reactions that are forbidden in flat space-time field theory. In this paper we want to propose a mechanism for generation of Z bosons in first order perturbative processes. Our study will be done by using the exact solutions of the Dirac equation and Proca equation written in the momentum helicity basis in an expanding background. This is an exact computation of a processes that implies production of massive bosons in early universe and the formalism could be adapted for computing other first order processes forbidden in Minkowski theory that generate bosons not only from vacuum, but also in emission processes by fermions.
For that, one needs to combine the General Relativity with the theory of electro-weak interactions. The problems that the Quantum Field Theory is facing in a curved spacetimes was the subject of many investigations. These investigations cover a vast area from the study of the free field equations in different metrics \cite{2,22,25,29,30} up to problems that imply interactions between fields and the renormalization theory. One of the subjects that receive attention is related to the mechanisms that generate the matter and antimatter and this subject was approached by perturbative methods\cite{15,18,23,24,26,27,28,31,b1,b2,36}, and nonperturbative methods \cite{13,14,16,17,32,33,34,35,38,39}.

We must also specify in our approach the modes are globally defined on entirely manifold and the vacuum state is stable and unique. In this case, the quantum states are measured using a global apparatus which consist of conserved operators \cite{24}. So the electro-weak transitions are measured by the same global apparatus which prepares all the quantum states, including the in and out asymptotic free
fields which remain minimally coupled to gravity. This means that the global apparatus cannot record particle creation
in the absence of the electromagnetic interaction \cite{24}, and we specify that the local detectors can record particle creation without electromagnetic interaction \cite{13,14}. Our approach does not contradict the cosmological particle creation which may be observed using local detectors. The results obtained by using perturbative methods can complete the general picture about the phenomenon of particle production, that was studied until now using predominantly non-perturbative methods.

Recent studies discussed the possible effects of adiabatic particle creations in the context of Loop Quantum Cosmology and with this approach the number of particles is computed by using thermodynamic arguments \cite{h}. Related to this result it will be of interest to translate all the perturbative results for probabilities in terms of number of particles, and a method for obtaining this result from a perturbative calculation was proposed in \cite{18}. Another method for study the production of particles was proposed in \cite{37}, and use the fact that the expansion of the universe give rise to a time-dependent gravitational field which generate particle production. Then a S matrix approach is used for compute the production probability for pairs to lowest order, by expanding the Lagrangian of general relativity with the help of metric tensor which is written in terms of flat space-time metric and a perturbation \cite{37}. The results are then compared with the corresponding pair creation probability by an external electromagnetic field \cite{37}.

In fact we study the effect of the electromagnetic interaction
upon the particle creation in the de Sitter expanding universe. The problem of production of massive Z bosons and neutrinos in the early universe by using various methods received little attention and was not approached by using perturbative methods. In this paper a method to study the generation of Z bosons and neutrinos from vacuum is proposed by computing the first order transition amplitudes corresponding to the neutral current interactions. In the Standard Model \cite{12,19,20} it is a well established fact that the first order transition amplitudes that generate Z bosons from vacuum are forbidden by the simultaneous conservation of energy and momentum. In a non-stationary metric this observation is no longer valid since the translational invariance with respect to time is lost and we specify that we work here in de Sitter geometry. The first step in our calculations will be to consider the formalism based on perturbative methods that allows to define the expression of the transition amplitudes in de Sitter spacetime. Then the generation from de Sitter vacuum of the triplet $Z$ boson,$\nu$ neutrino and $\tilde\nu$ antineutrino ($vac\rightarrow Z+\nu+\widetilde{\nu}$) will be analysed as a first order perturbative process. The time reversed process is also possible and represents the annihilation of the triplet in de Sitter vacuum. We mention that the expression for the transition amplitude and probability will be established by following the prescription from flat space-time perturbation theory in which the scattering operator is expressed in terms of interaction lagrangian density. The main steps for the computation of the transition amplitude and probability corresponding to the first order perturbative process will be presented and the physical consequences will be discussed. The computations are done in a de Sitter metric, and we use the chart that covers only the expanding part of the de Sitter variety.

In the second section we present the main steps for obtaining the definition for the transition amplitude in de Sitter geometry. The third section is dedicated to the computation of the transition probability for Z bosons and neutrinos generation from vacuum and we explore the physical consequences of the analytical results. In the forth section we present the main steps for computing the total probability and an analysis in terms of the ratio between mass of the Z boson and the expansion factor ($M_Z/\omega$) is done. In the fifth section we present our conclusions. We consider in our paper natural units such that $\hbar=1,c=1$.

\section{General formalism for obtaining the amplitude}
We start with the de Sitter metric written in conformal form \cite{1}:
\begin{equation}\label{metr}
ds^2=\frac{1}{(\omega t_{c})^2}(dt_{c}^2-d\vec{x}^2),
\end{equation}
where the conformal time is given in terms of proper time by $t_{c}=\frac{-e^{-\omega t}}{\omega}$ and $\omega$ is the expansion factor ($\omega>0$).  To define half-integer spin fields on curved spacetime one needs to use the tetrad fields \cite{2} $e_{\widehat{\mu}}(x)$ and $\widehat{e}^{\widehat{\mu}}(x)$,
which fix the local frames and corresponding coframes. These tetrad fields have local
indices $\widehat{\mu},\widehat{\nu},...=0,1,2,3$. For the
line element (\ref{metr}), the Cartesian gauge is chosen with
the nonvanishing tetrad components,
\begin{equation}
e^{0}_{\widehat{0}}=-\omega t_c  ;\,\,\,e^{i}_{\widehat{j}}=-\delta^{i}_{\widehat{j}}\,\omega t_c.
\end{equation}
Our study is done in the chart with conformal time $t_{c}\in(-\infty,0)$, which covers the expanding portion of de Sitter space.

By following the methods from flat space time the first step in constructing the theory of interactions between fields in de Sitter geometry will be to consider the free fields equations and their solutions. Then the free fields from $in$ and $out$ sectors are the exact solutions of the Proca equation and Dirac equation on de Sitter spacetime, written in the momentum helicity basis.
The transition amplitude for the spontaneous production from de Sitter vacuum of a $Z$ boson, a neutrino $(\nu)$ and antineutrino $(\tilde\nu)$ can be obtained by starting with the tetrad gauge invariant Lagrangian density that give the coupling between $Z$ bosons and leptons, written with point independent Dirac matrices $\gamma^{\hat\mu}$ and the tetrad fields $e_{\hat\mu}^{\alpha}$ :
\begin{eqnarray}\label{ll}
\mathcal{L}_{l\overline{l}Z}&=&-\left(\frac{e_0}{\sin(2\theta_{W})}\right)\overline{\psi}_{\nu_e}\,\gamma^{\hat\mu}e_{\hat\mu}^{\alpha}\left(\frac{1-\gamma^{5}}{2}\right)\psi_{\nu_e}A_{\alpha}(Z)
+\left(\frac{e_0\cos(2\theta_{W})}{\sin(2\theta_{W})}\right)\overline{\psi}_{e}\,\gamma^{\hat\mu}e_{\hat\mu}^{\alpha}\left(\frac{1-\gamma^{5}}{2}\right)\psi_e A_{\alpha}(Z)\nonumber\\
&&-(e_0\tan(\theta_{W}))\overline{\psi}_{e}\,\gamma^{\hat\mu}e_{\hat\mu}^{\alpha}\left(\frac{1+\gamma^{5}}{2}\right)\psi_e A_{\alpha}(Z)=-\frac{e_0}{\sin(2\theta_{W})}(j_{neutral})\,^{\alpha}A_{\alpha}(Z),
\end{eqnarray}
where $e_0$ is the electric charge, $\theta_{W}$ is the Weinberg angle and, $\psi_{\nu_{e}}$ designates the neutrino-antineutrino field, $A_{\alpha}(Z)$ designates the Z boson field and $\psi_e$ designates the electron-positron field. Since it is well known that the neutrinos are only left-handed we use the left projector $\frac{1-\gamma^{5}}{2}$, with the specification that for the electrons and positrons we have also the right-handed part and the corresponding right projector is $\frac{1+\gamma^{5}}{2}$. The Z boson have no electric charge and the particle coincides with the antiparticle, and mediates the  neutral  current interactions $(j_{neutral})\,^{\mu}$, and given by the following equation:
\begin{eqnarray}\label{l1}
(j_{neutral})\,^{\alpha}&=&\overline{\psi}_{\nu_e}\,\gamma^{\hat\mu}e_{\hat\mu}^{\alpha}\left(\frac{1-\gamma^{5}}{2}\right)\psi_{\nu_e}-\cos(2\theta_{W})\overline{\psi}_e\,\gamma^{\hat\mu}e_{\hat\mu}^{\alpha}\left(\frac{1-\gamma^{5}}{2}\right)\psi_e
\nonumber\\&&+2\sin^2(\theta_{W})\overline{\psi}_e\,\gamma^{\hat\mu}e_{\hat\mu}^{\alpha}\left(\frac{1+\gamma^{5}}{2}\right)\psi_e.
\end{eqnarray}
The above expression help us to establish the equation for the transition amplitudes by using perturbations.

Usually in the electro-weak theory \cite{3,4,5,6,7,8,9,10,11,12}, the amplitudes and probabilities are written by using the Feynmann rules in the momentum representation. Here we adopt for our computations the Feynmann rules in coordinates representation because our goal is to obtain the transition amplitude dependence on the expansion factor and in addition we do not have Feynmann rules in momentum picture since this will imply complex computations for internal lines of the graphs. Still we will present here the first steps for giving the Feynmann rules for the external lines of the graphs when massive Z boson interact with leptons.

It is known that the Proca equation and Dirac equation on de Sitter space-time can be analytically solved \cite{2,22}.
We begin with the free field that propagate in the $in$ and $out$ sectors. The solution for the Dirac equation in momentum-helicity basis which describe the zero mass particles with half integer spin on de Sitter spacetime was obtained in \cite{3}. Then the $U,V$ solutions that describe the neutrino $\nu$ and antineutrino $\tilde\nu$ in this geometry are \cite{22}:

\begin{eqnarray}\label{sol1}
(U_{\vec{p},\sigma}(x))_{\nu}=\left(-\frac{\omega t_{c}}{2\pi}\right)^{3/2}\left (\begin{array}{c}
(\frac{1}{2}-\sigma) \xi_{\sigma}(\vec{p}\,)\\
0
\end{array}\right)e^{i\vec{p}\cdot\vec{x}-i p t_{c}},\nonumber\\
(V_{\vec{p},\sigma}(x))_{\tilde\nu}=\left(-\frac{\omega t_{c}}{2\pi}\right)^{3/2}\left (\begin{array}{c}
(\frac{1}{2}+\sigma) \eta_{\sigma}(\vec{p}\,)\\
0
\end{array}\right)e^{-i\vec{p}\cdot\vec{x}+i p t_{c}}.
\end{eqnarray}
These solutions have only the left part (left-handed) and are obtained by using the chiral representation of Dirac matrices, such that the application of the left projector $\frac{1-\gamma^5}{2}$ leave them unchanged. The mode expansion in momentum representation is written in terms of the operators, $a,b$ and the
particle and antiparticle fundamental spinors $U,V$ which depend on the momentum
$p$ and polarization $\sigma=\pm\frac{1}{2}$:
\begin{eqnarray}
\psi(\vec{x},t)=\int d^3 p\sum_{\sigma}\left[a(\vec{p},\sigma)U_{\vec{p},\sigma}(x)+
b^+(\vec{p},\sigma)V_{\vec{p},\sigma}(x)\right].
\end{eqnarray}

The solutions of the Proca equation in de Sitter geometry written in the momentum helicity basis were obtained in \cite{2}. These solutions will describe the massive Z free field and their spatial part is:
\begin{eqnarray}\label{sol2}
\vec{f}_{\vec{\mathcal{P}},\lambda}(x)=\left\{
\begin{array}{cll}
\frac{i\sqrt{\pi}\omega \mathcal{P}e^{-\pi k/2}}{2M_Z(2\pi)^{3/2}}\left[(\frac{1}{2}+ik)\frac{\sqrt{-t_{c}}}{\mathcal{P}}
H^{(1)}_{ik}\left(-\mathcal{P}t_{c}\right)-(-t_{c})^{3/2}H^{(1)}_{1+ik}\left(-\mathcal{P}t_{c}\right)\right]e^{i\vec{\mathcal{P}}\vec{x}}\vec{\epsilon}\,(\vec{n}_{\mathcal{P}},\lambda)&{\rm for}&\lambda=0\\
\frac{\sqrt{\pi}e^{-\pi k/2}}{2(2\pi)^{3/2}}\sqrt{-t_{c}}H^{(1)}_{ik}\left(-\mathcal{P}t_{c}\right)e^{i\vec{\mathcal{P}}\vec{x}}\vec{\epsilon}\,(\vec{n}_{\mathcal{P}},\lambda)
&{\rm for}&\lambda=\pm 1.
\end{array}\right.
\nonumber\\
\end{eqnarray}
while the temporal component of the solution of Proca equation reads \cite{2}:
\begin{eqnarray}
f_{0\vec{\mathcal{P}},\lambda}(x)=\left\{
\begin{array}{cll}
\frac{\sqrt{\pi}\omega\mathcal{P}e^{-\pi k/2}}{2M_Z(2\pi)^{3/2}}(-t_{c})^{3/2}H^{(1)}_{ik}\left(-\mathcal{P}t_{c}\right)e^{i\vec{\mathcal{P}}\vec{x}}&{\rm for}&\lambda=0\\
0
&{\rm for}&\lambda=\pm 1.
\end{array}\right.
\end{eqnarray}

In the above equations for the plane wave for the Proca field $\vec{n}_{\mathcal{P}}=\vec{\mathcal{P}}/\mathcal{P}$ and $\vec{\epsilon}\,(\vec{n}_{\mathcal{P}},\lambda)$ are the polarization vectors. For $\lambda=\pm 1$ these vectors are transversal on the momentum such that
$\vec{\mathcal{P}}\cdot\vec{\epsilon}\,(\vec{n}_{\mathcal{P}},\lambda=\pm1)=0$ and for $\lambda=0$ the polarization vectors are longitudinal on the momentum
$\vec{\mathcal{P}}\cdot\vec{\epsilon}\,(\vec{n}_{\mathcal{P}},\lambda=0)=\mathcal{P}$, since $\vec{\epsilon}\,(\vec{n}_{\mathcal{P},\lambda=0})=\vec{n}_{\mathcal{P}}$. The mass of the Z boson is denoted by $M_Z$, while the parameter $k=\sqrt{\left(\frac{M_Z}{\omega}\right)^2-\frac{1}{4}}$ depends on the ratio $\frac{M_Z}{\omega}$, provided that  $\frac{M_Z}{\omega}>\frac{1}{2}$.

The mode expansion for Proca field is in momentum representation is written as:
\begin{eqnarray}
A^{\alpha}(\vec{x},t)=\int d^3\mathcal{P}\sum_{\lambda}\left[a(\vec{\mathcal{P}},\lambda)f_{\vec{\mathcal{P}},\lambda}^{\alpha}(x)+
a^+(\vec{\mathcal{P}},\lambda)f_{\vec{\mathcal{P}},\lambda}^{\alpha*}(x)\right],
\end{eqnarray}
where the operators $a(\vec{\mathcal{P}},\lambda)$ and fundamental solutions $f_{\vec{\mathcal{P}},\lambda}^{\alpha}(x)$ depend on momentum $\mathcal{P}$ and polarization $\lambda=0,\pm1$, and we mention that the field is real.

Because our study considers interactions with massive bosons, in computations both solutions for $\lambda=0$ and $\lambda=\pm1$ will be used, since the longitudinal modes also give contributions to the transition amplitude.

The Lagrangian density given in equation (\ref{l1}) can be used for establishing the general form of the transitions amplitudes when a Z boson interacts with neutrinos by using the perturbative methods as in flat space theory \cite{12,19,20}. Only the first term from equation (\ref{ll}) give contribution to our amplitude since in the $in/out$ sectors we have only neutrinos and a Z boson.
\begin{equation}
\mathcal{L}_{\nu\overline{\nu}Z}=-\left(\frac{e_0}{\sin(2\theta_{W})}\right)\overline{\psi}_{\nu_e}\,\gamma^{\hat\mu}e_{\hat\mu}^{\alpha}\left(\frac{1-\gamma^{5}}{2}\right)\psi_{\nu_e}A_{\alpha}(Z)
\end{equation}
The reduction rules for fermions are given in \cite{24}, while for the Z boson the reduction rules from the $in$ state and $out$ state can be computed using the method from flat space theory. In general the transition coefficients between two states can be computed as $\langle out;\alpha...|in;\beta...\rangle $
and represents the transition amplitudes between the $in$ states at $t\rightarrow-\infty$  and the $out$ states at $t\rightarrow\infty$.
Then the transition amplitudes can be constructed by using the reduction formalism. For the Proca field the reduction rules are:
\begin{eqnarray}
\langle out;\alpha|in; (\vec{\mathcal{P}},\lambda),\beta\rangle=\frac{i}{C}\int d^4x \sqrt{-g}\,\langle out;\alpha|A_\mu^+(x)|in;\beta\rangle \overleftarrow{E}_{P}(x)f_{\vec{\mathcal{P}},\lambda}^{\mu}(x),\nonumber\\
\langle out;(\vec{\mathcal{P}},\lambda),\alpha|in;\beta\rangle=\frac{i}{C}\int d^4x \sqrt{-g}\,f_{\vec{\mathcal{P}},\lambda}^{\mu*}(x)\overrightarrow{E}_{P}(x)\langle out;\alpha|A_\mu(x)|in;\beta\rangle,
\end{eqnarray}
where the notation ${E}_{P}(x)$ stands for the Proca operator ${E}_{P}(x)=\Box+m^2$ in de Sitter geometry \cite{2} and $C$ is a renormalization constant.
The perturbation calculations are based on the scattering operator given in terms of interaction lagrangean $\mathcal{L}_{l\overline{l}Z}$ as a expansion \cite{12,19,20}:
\begin{equation}
\textbf{S}=T \,\exp\left[-i\int d^4x \sqrt{-g}\mathcal{L}_{l\overline{l}Z}\right].
\end{equation}

Then by using the scattering operator we can compute the Green functions in terms of free fields:
\begin{equation}
\langle 0|\Psi(x_1)\bar{\Psi}(x_2) A_{\mu}(x_3)...|0\rangle=\frac{\langle0|\Psi(x_1)\bar{\Psi}(x_2) A_{\mu}(x_3)...\textbf{S}|0\rangle}{\langle0|\textbf{S}|0\rangle}.
\end{equation}
The above Green functions can be expressed in terms of Feynman propagators
if we consider all the possible T contractions as the
Wick theorem states. Then the numerators of these functions
can be split in connected parts multiplied just by the
vacuum expectation value $\langle0|\textbf{S}|0\rangle$ such that after simplification
we remain only with the connected parts witch give the transition amplitudes. The transition amplitude in the first order of the perturbation theory for the interaction between Z boson and neutrino-antineutrino field reads:
\begin{eqnarray}\label{aper}
&&\mathcal{A}_{Z\nu\overline{\nu}}=-\int d^4x\sqrt{-g}\,\left(\frac{e_0}{\sin(2\theta_{W})}\right)\overline{\psi}_{\nu_e}\,\gamma^{\hat\mu}e_{\hat\mu}^{\alpha}\left(\frac{1-\gamma^{5}}{2}\right)\psi_{\nu_{e}}A_{\alpha}(Z).
\end{eqnarray}
By using the same method the first order transition amplitude that describe the interaction between Z boson and electron-positron field is
\begin{eqnarray}
\mathcal{A}_{Ze\overline{e}}&=&\int d^4x\sqrt{-g}\,\biggl\{\left(\frac{e_0\cos(2\theta_{W})}{\sin(2\theta_{W})}\right)\overline{\psi}_e\,\gamma^{\hat\mu}e_{\hat\mu}^{\alpha}\left(\frac{1-\gamma^{5}}{2}\right)\psi_e A_{\alpha}(Z)\nonumber\\
&&-(e_0\tan(\theta_{W}))\overline{\psi}_e\,\gamma^{\hat\mu}e_{\hat\mu}^{\alpha}\left(\frac{1+\gamma^{5}}{2}\right)\psi_eA_{\alpha}(Z)\biggl\}.
\nonumber\\
\end{eqnarray}
The above equations can be used for study all the interactions between Z bosons and leptons in the first order of the perturbation theory, in a de Sitter geometry.

\section{Amplitude and probability computation}
This section is dedicated to amplitude and probability computation for the process of spontaneous generation from de Sitter vacuum of the triplet Z boson, neutrino and antineutrino. The amplitude of the process can be computed by using equation (\ref{aper}), in which we replace
the solutions of the Dirac equation and Proca equation in de Sitter spacetime.
Then the amplitude equation (\ref{aper}) can be expanded using the temporal part solution $f_{0\mathcal{P},\lambda}(x)$ and spatial part solution  $f_{j\,\mathcal{P},\lambda}(x)$ of the Proca equation \cite{2} as:
\begin{eqnarray}\label{ampltt}
\mathcal{A}_{Z\nu\overline{\nu}}=-\int d^4x\sqrt{-g}\left(\frac{e_0}{\sin(2\theta_{W})}\right)(\overline{U}_{p,\sigma})_{\nu}(x)\gamma^{\hat\mu}e_{\,\hat\mu}^{\alpha}\left(\frac{1-\gamma^{5}}{2}\right)
(V_{p'\sigma'})_{\tilde\nu}(x)f_{\alpha\mathcal{P},\lambda,Z}^*(x)\nonumber\\
=-\int d^4x\sqrt{-g}\left(\frac{e_0}{\sin(2\theta_{W})}\right)(\overline{U}_{p,\sigma})_{\nu}(x)\gamma^{\hat0}e_{\,\hat0}^{0}\left(\frac{1-\gamma^{5}}{2}\right)
(V_{p'\sigma'})_{\tilde\nu}(x)f_{0\mathcal{P},\lambda,Z}^*(x)\nonumber\\
-\int d^4x\sqrt{-g}\left(\frac{e_0}{\sin(2\theta_{W})}\right)(\overline{U}_{p,\sigma})_{\nu}(x)\gamma^{\hat i}e_{\,\hat i}^{j}\left(\frac{1-\gamma^{5}}{2}\right)
(V_{p'\sigma'})_{\tilde\nu}(x)f_{j\,\mathcal{P},\lambda,Z}^*(x).
\end{eqnarray}
\subsection{The calculation}
The equation (\ref{ampltt}) will be used for computing the amplitude for $\lambda=0$ , and in this case we have a continuous dependence on temporal solution and spatial solution of the Proca equation in de Sitter space-time:
\begin{eqnarray}\label{am0}
\mathcal{A}_{Z\nu\overline{\nu}}(\lambda=0)=-\int d^4x\sqrt{-g}\left(\frac{e_0}{\sin(2\theta_{W})}\right)(\overline{U}_{p,\sigma})_{\nu}(x)\gamma^{\hat0}e_{\,\hat0}^{0}\left(\frac{1-\gamma^{5}}{2}\right)
(V_{p'\sigma'})_{\tilde\nu}(x)f_{0\mathcal{P},\lambda=0,Z}^*(x)\nonumber\\
-\int d^4x\sqrt{-g}\left(\frac{e_0}{\sin(2\theta_{W})}\right)(\overline{U}_{p,\sigma})_{\nu}(x)\gamma^{\hat i}e_{\,\hat i}^{j}\left(\frac{1-\gamma^{5}}{2}\right)
(V_{p'\sigma'})_{\tilde\nu}(x)f_{j\,\mathcal{P},\lambda=0,Z}^*(x).
\end{eqnarray}
In the case $\lambda=\pm1$ only the spatial part of the solution give contributions since the temporal part is zero and we obtain:
\begin{eqnarray}\label{am1}
\mathcal{A}_{Z\nu\overline{\nu}}(\lambda=\pm1)=-\int d^4x\sqrt{-g}\left(\frac{e_0}{\sin(2\theta_{W})}\right)(\overline{U}_{p,\sigma})_{\nu}(x)\gamma^{\hat i}e_{\,\hat i}^{j}\left(\frac{1-\gamma^{5}}{2}\right)
(V_{p'\sigma'})_{\tilde\nu}(x)f_{j\,\mathcal{P},\lambda=\pm1,Z}^*(x).
\nonumber\\
\end{eqnarray}
The amplitude equations (\ref{am0}) and (\ref{am1}), can be split in temporal integrals and spatial integrals. The spatial integral contain the delta Dirac function expressing the momentum conservation in this process, while for the temporal integral the new integration variable $z=-t_{c}$ \cite{23,24,26} is introduced, and the result for the above amplitudes is:
\begin{eqnarray}
&&\mathcal{A}_{Z\nu\overline{\nu}}(\lambda=0)=\frac{e_0}{\sin(2\theta_{W})}\,\delta^3(\vec{\mathcal{P}}+\vec{p}+\vec{p}\,')\frac{1}{\sqrt{\pi}(2\pi)^{3/2}}
\left(\frac{1}{2}-\sigma\right)\left(\frac{1}{2}+\sigma'\right)\nonumber\\
&&\times\biggl\{ \frac{\mathcal{P}\omega}{M_Z}A(t_{c})\xi^+_{\sigma}(\vec{p}\,)\vec{\sigma}\cdot\vec{\epsilon}\,^*(\vec{n}_{\mathcal{P}},\lambda=0)\eta_{\sigma'}(\vec{p}\,')
+\frac{\mathcal{P}\omega}{M_Z}C(t_{c})\xi^+_{\sigma}(\vec{p}\,)\eta_{\sigma'}(\vec{p}\,')\biggl\},
\end{eqnarray}
\begin{eqnarray}
&&\mathcal{A}_{Z\nu\overline{\nu}}(\lambda=\pm1)=\frac{e_0}{\sin(2\theta_{W})}\,\delta^3(\vec{\mathcal{P}}+\vec{p}+\vec{p}\,')\frac{1}{\sqrt{\pi}(2\pi)^{3/2}}
\left(\frac{1}{2}-\sigma\right)\left(\frac{1}{2}+\sigma'\right)\nonumber\\
&&\times\biggl\{B(t_{c})\xi^+_{\sigma}(\vec{p}\,)\vec{\sigma}\cdot\vec{\epsilon}\,^*(\vec{n}_{\mathcal{P}},\lambda=\pm1)\eta_{\sigma'}(\vec{p}\,')\biggl\},
\end{eqnarray}
where $A(t_{c}),B(t_{c}),C(t_c)$ are the temporal integrals given by:
\begin{eqnarray}
A(t_{c})&=&\int_0^\infty dz \sqrt{z}\,e^{-i(p+p')z}K_{-ik}(i\mathcal{P}z)\frac{1}{\mathcal{P}}\left(\frac{1}{2}-ik\right)-i\int_0^\infty dz z^{3/2}\,e^{-i(p+p')z}K_{1-ik}(i\mathcal{P}z),\nonumber\\
C(t_c)&=&i\int_0^\infty dz z^{3/2}\,e^{-i(p+p')z}K_{-ik}(i\mathcal{P}z)\nonumber\\
B(t_{c})&=&i\int_0^\infty dz \sqrt{z}\,e^{-i(p+p')z}K_{-ik}(i\mathcal{P}z).
\end{eqnarray}
The results of the integrals are discussed in equation (\ref{a3}) from Appendix . Taking into account all contributions the final result for the transition amplitudes is:
\begin{eqnarray}
&&\mathcal{A}_{Z\nu\overline{\nu}}(\lambda=0)=\frac{e_0}{\sin(2\theta_{W})}\,\delta^3(\vec{\mathcal{P}}+\vec{p}+\vec{p}\,\,')\frac{1}{(2\pi)^{3/2}}
\left(\frac{1}{2}-\sigma\right)\left(\frac{1}{2}+\sigma'\right)\nonumber\\
&&\times\biggl\{A_{k}(\mathcal{P},p,p')\xi^+_{\sigma}(\vec{p}\,)\vec{\sigma}\cdot\vec{\epsilon}\,^*(\vec{n}_{\mathcal{P}},\lambda=0)\eta_{\sigma'}(\vec{p}\,')\nonumber\\
&&+C_{k}(\mathcal{P},p,p')\xi^+_{\sigma}(\vec{p}\,)\eta_{\sigma'}(\vec{p}\,')\biggl\},\nonumber\\
&&\mathcal{A}_{Z\nu\overline{\nu}}(\lambda=\pm1)=\frac{e_0}{\sin(2\theta_{W})}\,\delta^3(\vec{\mathcal{P}}+\vec{p}+\vec{p}\,\,')\frac{1}{(2\pi)^{3/2}}
\left(\frac{1}{2}-\sigma\right)\left(\frac{1}{2}+\sigma'\right)\nonumber\\
&&\times\biggl\{B_{k}(\mathcal{P},p,p')\xi^+_{\sigma}(\vec{p}\,)\vec{\sigma}\cdot\vec{\epsilon}\,^*(\vec{n}_{\mathcal{P}},\,\lambda=\pm1)\eta_{\sigma'}(\vec{p}\,')\biggl\}
\end{eqnarray}
In the above equations we introduced the following notations:
\begin{eqnarray}\label{ab}
&&A_{k}(\mathcal{P},p,p')=\frac{i^{-3/2}(2\mathcal{P})^{-ik}}{(\mathcal{P}+p+p')^{3/2-ik}}\frac{\omega}{M_Z}
\Gamma\left(\frac{3}{2}-ik\right)\Gamma\left(\frac{3}{2}+ik\right)\left(\frac{1}{2}-ik\right)\nonumber\\
&&\times_{2}F_{1}\left(\frac{3}{2}-i
k,\frac{1}{2}-ik;2;\frac{-\mathcal{P}+p+p'}{\mathcal{P}+p+p\,'}\right)
-\frac{i^{-3/2}\mathcal{P}(2\mathcal{P})^{1-ik}}{2(\mathcal{P}+p+p')^{7/2-ik}}\frac{\omega}{M_Z}
\Gamma\left(\frac{7}{2}-ik\right)\nonumber\\
&&\times\Gamma\left(\frac{3}{2}+ik\right)\,_{2}F_{1}\left(\frac{7}{2}-i
k,\frac{3}{2}-ik;3;\frac{-\mathcal{P}+p+p'}{\mathcal{P}+p+p\,'}\right),\nonumber\\
&&C_k(\mathcal{P},p,p')=\frac{i^{-3/2}\mathcal{P}(2\mathcal{P})^{-ik}}{2(\mathcal{P}+p+p')^{5/2-ik}}\frac{\omega}{M_Z}
\Gamma\left(\frac{5}{2}-ik\right)\Gamma\left(\frac{5}{2}+ik\right)\nonumber\\
&&\times\,_{2}F_{1}\left(\frac{5}{2}-i
k,\frac{1}{2}-ik;3;\frac{-\mathcal{P}+p+p'}{\mathcal{P}+p+p\,'}\right),
\end{eqnarray}
\begin{eqnarray}\label{cb}
&&B_{k}(\mathcal{P},p,p')=\frac{i^{-1/2}(2\mathcal{P})^{-ik}}{(\mathcal{P}+p+p')^{3/2-ik}}
\Gamma\left(\frac{3}{2}-ik\right)\Gamma\left(\frac{3}{2}+ik\right)\nonumber\\
&&\times\,_{2}F_{1}\left(\frac{3}{2}-i
k,\frac{1}{2}-ik;2;\frac{-\mathcal{P}+p+p'}{\mathcal{P}+p+p\,'}\right),
\nonumber\\
\end{eqnarray}

The delta Dirac terms $\delta^3(\vec{\mathcal{P}}+\vec{p}+\vec{p}\,\,')$ assures the momentum conservation in the process of Z boson and neutrinos spontaneous generation from vacuum. The result related to the momentum conservation was also obtained in de Sitter QED, when the first order processes that generate triplets from vacuum were studied \cite{24}.
The analytical structure of the final result is depending on Gauss hypergeometric functions $_{2}F_{1}$ and gamma Euler functions $\Gamma$. Their dependence of gravity enters in equation (\ref{ab}) via the parameter $k=\sqrt{\left(\frac{M_Z}{\omega}\right)^2-\frac{1}{4}}$. Remarkably is that the ratio between Z boson mass and the expansion parameter $\frac{M_Z}{\omega}$ and the momenta $p,p',\mathcal{P}$ completely determine the amplitude and probability. This suggest an analysis implying these parameters, looking for the behaviour of the functions $A_k,B_k,C_k$ in terms of the parameter $M_Z/\omega$.

The probability is obtained by summing after the final helicities the square modulus of the amplitude. Here a few observations are in order, first we observe that the amplitude is nonvanishing only for specific values of fermion helicities, $\sigma=-\frac{1}{2}, \sigma'=\frac{1}{2}$ and it is no longer need to sum over them. More precisely there are selection rules such that if the helicity of the fermion is $\sigma$ then the anti-fermion helicity will always be $\sigma'=-\sigma$.
Since the term with the delta Dirac function $\delta^3(\vec{\mathcal{P}}+\vec{p}+\vec{p}\,\,')$ is present in amplitude, we will define here the transition probability per volume unit , i.e. $|\delta^3(\vec{p}\,)|^2=V\delta^3(\vec{p}\,)$ :
\begin{eqnarray}\label{pif}
&&P_{Z\nu\overline{\nu}}(\lambda=0)=|\mathcal{A}_{Z\nu\overline{\nu}}(\lambda=0)|^2=
\frac{e_0^2}{\sin^2(2\theta_{W})}\,\delta^3(\vec{\mathcal{P}}+\vec{p}+\vec{p}\,\,')\frac{1}{(2\pi)^{3}}\nonumber\\
&&\left(\frac{1}{2}-\sigma\right)^2\left(\frac{1}{2}+\sigma'\right)^2\biggl\{
|A_{k}(\mathcal{P},p,p')|^2|\xi^+_{\sigma}(\vec{p}\,)\vec{\sigma}\cdot\vec{\epsilon}\,^*(\vec{n}_{\mathcal{P}},\lambda=0)\eta_{\sigma'}(\vec{p}\,')|^2\nonumber\\
&&+|C_{k}(\mathcal{P},p,p')|^2|\xi^+_{\sigma}(\vec{p}\,)\eta_{\sigma'}(\vec{p}\,')|^2\nonumber\\
&&+A_{k}^*(\mathcal{P},p,p')C_{k}(\mathcal{P},p,p')(\xi^+_{\sigma}(\vec{p}\,)\vec{\sigma}\cdot\vec{\epsilon}\,^*(\vec{n}_{\mathcal{P}},\lambda=0)\eta_{\sigma'}(\vec{p}\,'))^*(\xi^+_{\sigma}(\vec{p}\,)\eta_{\sigma'}(\vec{p}\,'))\nonumber\\
&&+C_{k}^*(\mathcal{P},p,p')A_{k}(\mathcal{P},p,p')(\xi^+_{\sigma}(\vec{p}\,)\eta_{\sigma'}(\vec{p}\,'))^*(\xi^+_{\sigma}(\vec{p}\,)\vec{\sigma}\cdot\vec{\epsilon}\,^*(\vec{n}_{\mathcal{P}},\lambda=0)\eta_{\sigma'}(\vec{p}\,'))\biggl\}
\end{eqnarray}
\begin{eqnarray}\label{pif1}
&&P_{Z\nu\overline{\nu}}(\lambda=\pm1)=\frac{1}{2}\sum_{\lambda}|\mathcal{A}_{Z\nu\overline{\nu}}(\lambda=\pm1)|^2=\frac{e_0^2}{\sin^2(2\theta_{W})}\,\delta^3(\vec{\mathcal{P}}+\vec{p}+\vec{p}\,\,')\frac{1}{(2\pi)^{3}}\nonumber\\
&&\left(\frac{1}{2}-\sigma\right)^2\left(\frac{1}{2}+\sigma'\right)^2\biggl\{\frac{1}{2}\sum_{\lambda}|B_{k}(\mathcal{P},p,p')|^2|\xi^+_{\sigma}(\vec{p}\,)\vec{\sigma}\cdot\vec{\epsilon}\,^*(\vec{n}_{\mathcal{P}},\,\lambda=\pm1)\eta_{\sigma'}(\vec{p}\,')|^2\biggl\}.
\end{eqnarray}
We observe from the probability equations that there are two contributions in probability coming from longitudinal modes with $\lambda=0$ and the  transversal modes with $\lambda=\pm 1$.

\subsection{Graphical analysis}
The key parameters in our amplitude are the momenta of the produced particles $p,p',\mathcal{P}$ and the ratio between $Z$ boson mass and the expansion parameter, $M_Z/\omega$. Since all the functions that define the amplitude and probability depend on the parameter $k=\sqrt{\left(\frac{M_Z}{\omega}\right)^2-\frac{1}{4}}$, we must take into account that and the condition in which our computations are done is $\frac{M_Z}{\omega}>\frac{1}{2}$ such that the index of the Hankel functions are imaginary. In these conditions our graphs were done for the interval $\frac{M_Z}{\omega}\in(\frac{1}{2},\infty]$.
\begin{figure}[h!t]
\includegraphics[scale=0.4]{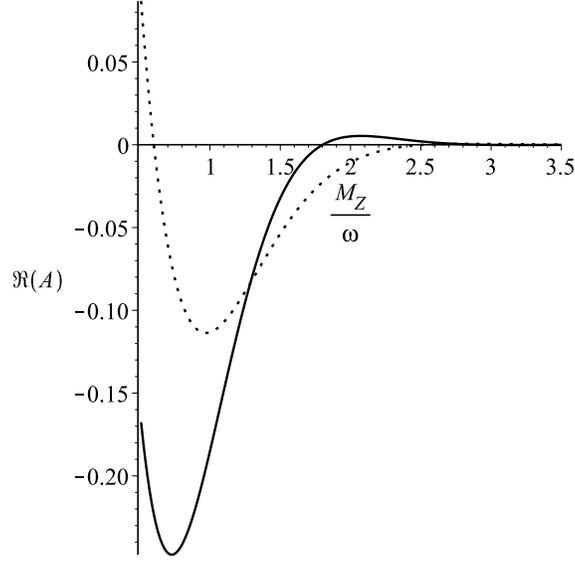}
\caption{Real part of $A_{k}$ as a function of parameter $M_Z/\omega$. Solid line is for $p=0.5,p'=0.4,\mathcal{P}=0.1$, while the point line is for $p=0.5,p'=0.3,\mathcal{P}=0.2$ .}
\label{f1}
\end{figure}
\begin{figure}[h!t]
\includegraphics[scale=0.4]{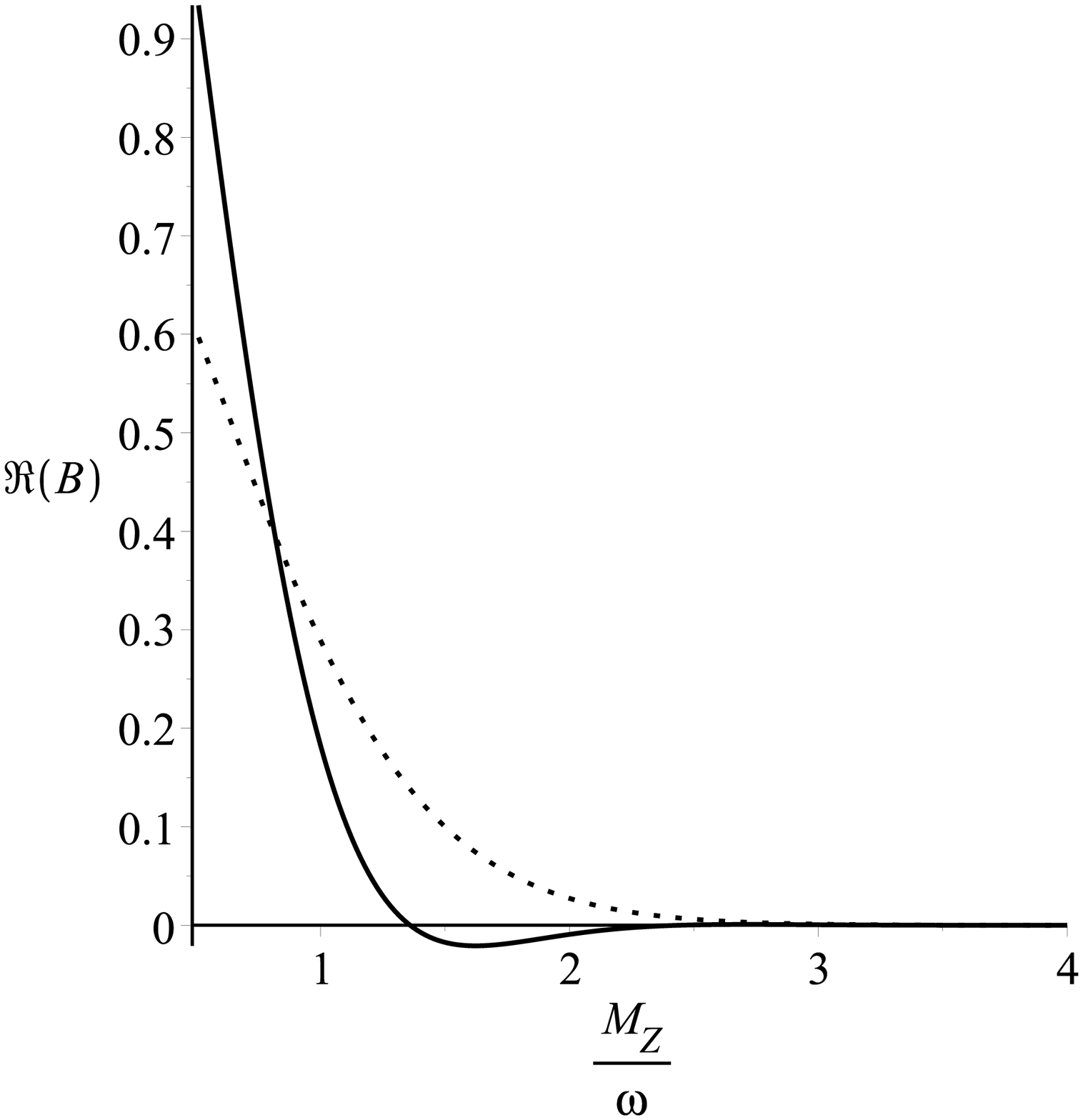}
\caption{Real part of $B_{k}$ as a function of parameter $M_Z/\omega$. Solid line is for $p=0.5,p'=0.4,\mathcal{P}=0.1$, while the point line is for $p=0.2,p'=0.4,\mathcal{P}=0.4$. }
\label{f2}
\end{figure}
\begin{figure}[h!t]
\includegraphics[scale=0.4]{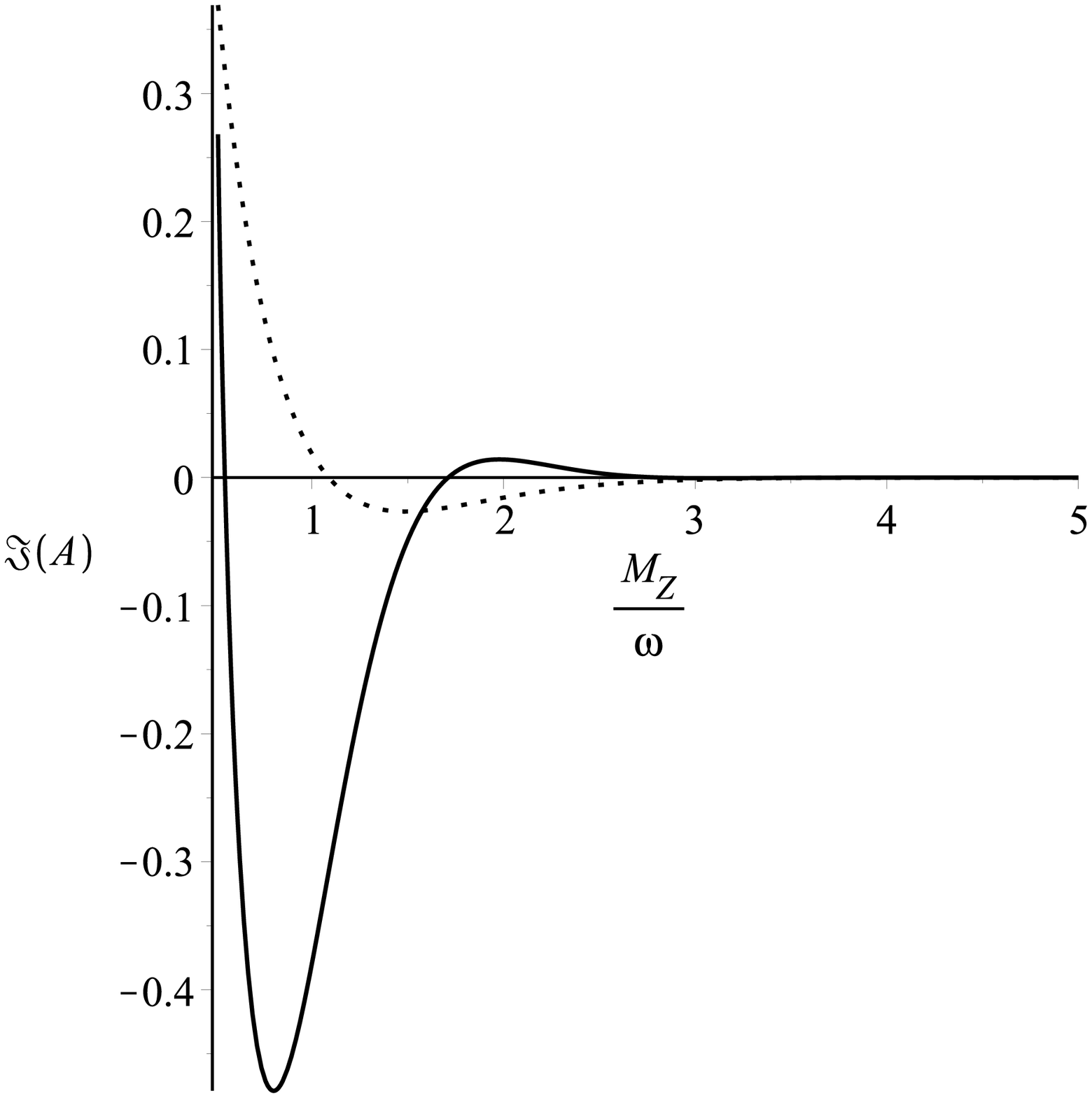}
\caption{Imaginary part of $A_{k}$ as a function of parameter $M_Z/\omega$. Solid line is for $p=0.4,p'=0.5,\mathcal{P}=0.1$, while the point line is for $p=0.2,p'=0.4,\mathcal{P}=0.4$. }
\label{f3}
\end{figure}
\begin{figure}[h!t]
\includegraphics[scale=0.4]{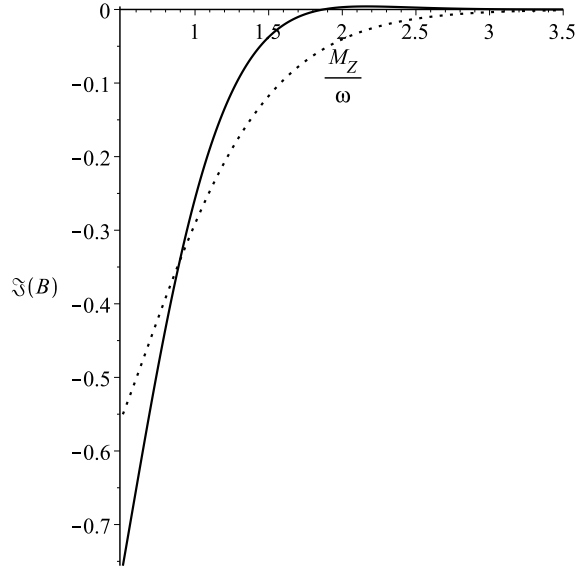}
\caption{Imaginary part of $B_{k}$ as a function of parameter $M_Z/\omega$. Solid line is for $p=0.3,p'=0.5,\mathcal{P}=0.2$, while the point line is for $p=0.2,p'=0.3,\mathcal{P}=0.5$.}
\label{f4}
\end{figure}
The results presented in figs.(\ref{f1})-(\ref{f4}) prove that the functions $A_{k},B_{k}$ which define the amplitude converge in both their real and imaginary parts, if the analysis is done in terms of the parameter $M_Z/\omega$, and fixed values for the momenta $p,p',\mathcal{P}$. Another observation is that for $M_Z/\omega\rightarrow\infty$, the Minkowski limit can be discussed since this corresponds to $\omega\rightarrow0$ and we observe from figs. (\ref{f1})-(\ref{f4}) that the functions $A_{k},B_{k}$ are vanishing in this limit. The same observations are valid for the function $C_k$.

The probability can be understood better by plotting the square modulus of the functions $A_{k},B_{k},C_k$ in terms of $M_Z/\omega$.
\begin{figure}[h!t]
\includegraphics[scale=0.4]{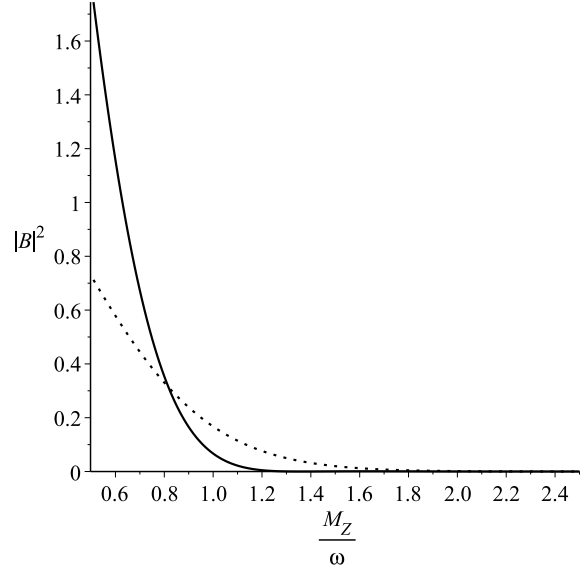}
\caption{ $|B_{k}|^2$ as a function of parameter $M_Z/\omega$. Solid line is for $p=0.5,p'=0.4,\mathcal{P}=0.1$, while the point line is for $p=0.5,p'=0.3,\mathcal{P}=0.2$ .}
\label{f5}
\end{figure}
\begin{figure}[h!t]
\includegraphics[scale=0.4]{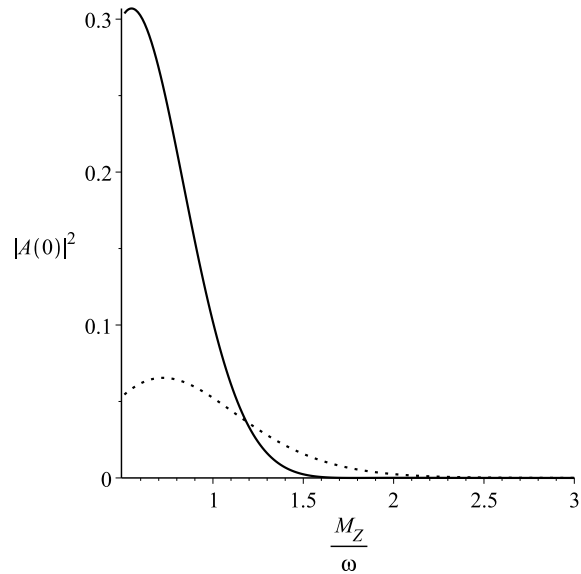}
\caption{$|A(0)|^2$ as a function of parameter $M_Z/\omega$. Solid line is for $p=0.5,p'=0.4,\mathcal{P}=0.1$, while the point line is for $p=0.5,p'=0.3,\mathcal{P}=0.2$.}
\label{f6}
\end{figure}.

\begin{figure}[h!t]
\includegraphics[scale=0.4]{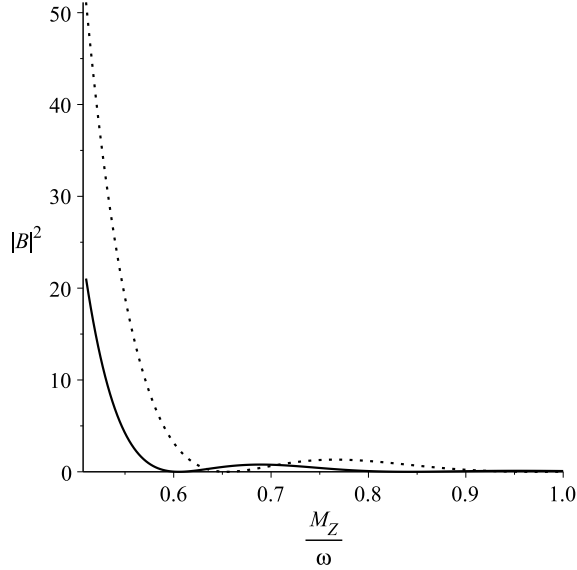}
\caption{$|B_{k}|^2$ as a function of parameter $M_Z/\omega$. Solid line is for $p=0.5,p'=0.4,\mathcal{P}=0.0001$, while the point line is for $p=0.2,p'=0.4,\mathcal{P}=0.0004$. }
\label{f7}
\end{figure}

\begin{figure}[h!t]
\includegraphics[scale=0.4]{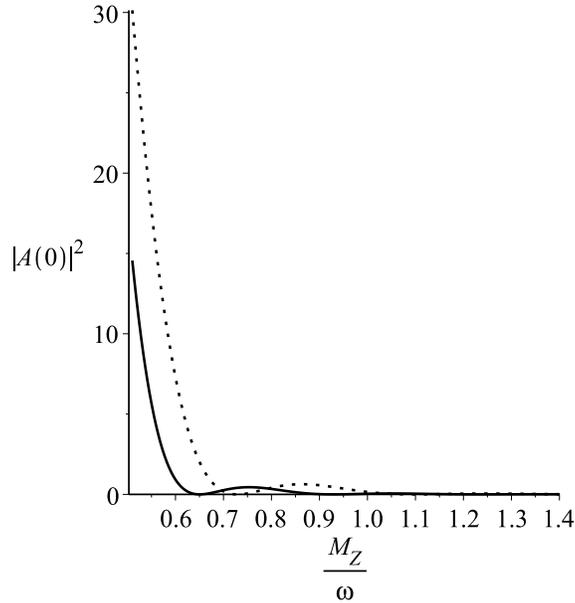}
\caption{$|A(0)|^2$ as a function of parameter $M_Z/\omega$. Solid line is for $p=0.5,p'=0.4,\mathcal{P}=0.0001$, while the point line is for $p=0.4,p'=0.2,\mathcal{P}=0.0004$.}
\label{f8}
\end{figure}

An interesting observation can be done considering that the momenta of the Z boson $\mathcal{P}$ is much more smaller that the momenta of the neutrino-antineutrino pair $p,p'$, and then plot the probability for the transversal modes $\lambda=\pm 1$, and longitudinal modes $\lambda=0$. Then in this case the probability shows a oscillatory behaviour, which is contained in the same interval where the ratio $M_Z/\omega$ is small (see figs.(\ref{f7})-(\ref{f8})).

In fact the quantity computed is the probability density which integrated after the final momenta will give the total probability and this analysis is the topic of the next section. From the last two graphs fig.((\ref{f7})-(\ref{f8})), we observe that considering very small values for the Z boson momenta as modulus, then the probability is significatively larger than in the case when the momenta of the Z boson is comparable with the momenta of the neutrino and anti-neutrino in modulus. We conclude that emission of "soft" Z bosons is the favoured process under space expansion.

The above graphs fig.(\ref{f5})-(\ref{f8}) for the $|A(0)|^2, |B_{k}|^2$ give the behaviour of the probability in terms of parameter $M_Z/\omega$, the factor $|A(0)|^2=|A_{k}|^2+|C_{k}|^2+A_{k}^*C_k+A_{k}C_k^*$ contain all the contributions from probability in the case $\lambda=0$. The results of the graphical analysis prove that the process of particle production from de Sitter vacuum is important only in strong gravitational fields $\omega\geq M_Z$. Our result prove that the production of Z boson is possible only in early universe when the expansion parameter was larger as the Z boson mass or have values comparable with the Z boson mass. This result is remarkable since it is the first computation that prove that Z bosons could be produced in strong gravitational fields as a first order perturbative process, that is forbidden in the Minkowski electro-weak theory \cite{10,11,12}. Our study suggest that a more general theory could be developed in de Sitter space-time, and that from this theory one should recover the well known results from Minkowski theory in the limit of zero expansion parameter.

\subsection{Helicity bispinor summation}

Further computation of the terms dependent on helicity spinors and polarization vectors could be done by taking an orthogonal local frame defined by the basis vectors $\vec{e}_i$. In this local frame define the momenta $\vec{p}=p_i\vec{e}_i$ and for the beginning fix the Z boson momenta on third axis such that $\vec{\mathcal{P}}=\mathcal{P}\vec{e}_3$. For the momenta of the neutrinos we will take the spherical coordinates such that they move in the plane $(1,3)$ i.e. $\vec{p}\,(p,\alpha,\beta=0)$ and $\vec{p}\,'(p',\gamma,\theta=\pi)$ \cite{24}. In this setup the angle between $\vec{p}$ and $\vec{p}\,'$ is just $\alpha+\gamma$. The momenta conservation in this process, when projected on the two axes that define the plane $(1,3)$ give:
\begin{eqnarray}
\mathcal{P}=p\cos\alpha+p'\cos\gamma\,\,; \nonumber\\
p\sin\alpha-p'\sin\gamma=0.
\end{eqnarray}
From these equations we deduce that:
\begin{equation}
\frac{p}{\mathcal{P}}=\frac{\sin\gamma}{\sin(\alpha+\gamma)},\,\frac{p'}{\mathcal{P}}=\frac{\sin\alpha}{\sin(\alpha+\gamma)}
\end{equation}
The functions $A_{k},B_{k},C_{k}$ that define the amplitude and contain momenta dependence can be expressed using the above relations, and finally obtain:

\begin{eqnarray}\label{abc}
&&A_{k}(\mathcal{P},p,p')=\frac{i^{-3/2}(2)^{-ik}}{\mathcal{P}^{3/2}(1+\frac{\sin\gamma}{\sin(\alpha+\gamma)}+\frac{\sin\alpha}{\sin(\alpha+\gamma)})^{3/2-ik}}\frac{\omega}{M_Z}
\Gamma\left(\frac{3}{2}-ik\right)\Gamma\left(\frac{3}{2}+ik\right)\left(\frac{1}{2}-ik\right)\nonumber\\
&&\times_{2}F_{1}\left(\frac{3}{2}-i
k,\frac{1}{2}-ik;2;\frac{-1+\frac{\sin\gamma}{\sin(\alpha+\gamma)}+\frac{\sin\alpha}{\sin(\alpha+\gamma)}}{1+\frac{\sin\gamma}{\sin(\alpha+\gamma)}+\frac{\sin\alpha}{\sin(\alpha+\gamma)}}\right)
-\frac{i^{-3/2}(2)^{1-ik}}{2\mathcal{P}^{3/2}(1+\frac{\sin\gamma}{\sin(\alpha+\gamma)}+\frac{\sin\alpha}{\sin(\alpha+\gamma)})^{7/2-ik}}\nonumber\\
&&\times\frac{\omega}{M_Z}\Gamma\left(\frac{7}{2}-ik\right)\Gamma\left(\frac{3}{2}+ik\right)\,_{2}F_{1}\left(\frac{7}{2}-i
k,\frac{3}{2}-ik;3;\frac{-1+\frac{\sin\gamma}{\sin(\alpha+\gamma)}+\frac{\sin\alpha}{\sin(\alpha+\gamma)}}{1+\frac{\sin\gamma}{\sin(\alpha+\gamma)}+\frac{\sin\alpha}{\sin(\alpha+\gamma)}}\right),\nonumber\\
&&C_k(\mathcal{P},p,p')=\frac{i^{-3/2}(2)^{-ik}}{2\mathcal{P}^{3/2}(1+\frac{\sin\gamma}{\sin(\alpha+\gamma)}+\frac{\sin\alpha}{\sin(\alpha+\gamma)})^{5/2-ik}}\frac{\omega}{M_Z}\Gamma\left(\frac{5}{2}-ik\right)\Gamma\left(\frac{5}{2}+ik\right)\nonumber\\
&&\times\,_{2}F_{1}\left(\frac{5}{2}-i
k,\frac{1}{2}-ik;3;\frac{-1+\frac{\sin\gamma}{\sin(\alpha+\gamma)}+\frac{\sin\alpha}{\sin(\alpha+\gamma)}}{1+\frac{\sin\gamma}{\sin(\alpha+\gamma)}+\frac{\sin\alpha}{\sin(\alpha+\gamma)}}\right),\nonumber\\
&&B_{k}(\mathcal{P},p,p')=\frac{i^{-1/2}(2)^{-ik}}{\mathcal{P}^{3/2}(1+\frac{\sin\gamma}{\sin(\alpha+\gamma)}+\frac{\sin\alpha}{\sin(\alpha+\gamma)})^{3/2-ik}}
\Gamma\left(\frac{3}{2}-ik\right)\Gamma\left(\frac{3}{2}+ik\right)\nonumber\\
&&\times\,_{2}F_{1}\left(\frac{3}{2}-i
k,\frac{1}{2}-ik;2;\frac{-1+\frac{\sin\gamma}{\sin(\alpha+\gamma)}+\frac{\sin\alpha}{\sin(\alpha+\gamma)}}{1+\frac{\sin\gamma}{\sin(\alpha+\gamma)}+\frac{\sin\alpha}{\sin(\alpha+\gamma)}}\right).
\nonumber\\
\end{eqnarray}

The above formula depending on the angle between momenta vectors $\vec{p},\vec{p}\,'$ help us to establish the probability equation for different momenta configurations. The last step in completing the probability equation is to compute the helicity bispinor products. First we analyse the situation when $\lambda=0$, and considering the above case with $\vec{\mathcal{P}}=\mathcal{P}\vec{e}_3$ and knowing that  $\vec{\mathcal{P}}\cdot\vec{\epsilon}\,(\vec{n}_{\mathcal{P}},\lambda=0)=\mathcal{P}$, then $\vec{\epsilon}\,(\vec{n}_{\mathcal{P}},\lambda=0)=\vec{e}_3$. Now taking into account that fermion helicities take only the values $\sigma=-\frac{1}{2},\,\sigma'=\frac{1}{2}$ we finally obtain the form of the helicity bispinors by using equation (\ref{spin}) from Appendix:
\begin{eqnarray}\label{spins}
&&\xi_{-\frac{1}{2}}(\vec{p}\,)=\left(
  \begin{array}{c}
    -\sin\left(\frac{\alpha}{2}\right) \\
   \cos\left(\frac{\alpha}{2}\right) \\
  \end{array}
\right)\nonumber\\
&&\eta_{\frac{1}{2}}(\vec{p}\,)=\left(
  \begin{array}{c}
    -\sin\left(\frac{\gamma}{2}\right) \\
    -\cos\left(\frac{\gamma}{2}\right) \\
  \end{array}
\right).
\end{eqnarray}
The last equation, help us to compute the terms dependent on helicity bispinors from our amplitude and probability. Then for $\lambda=0$ the results are:
\begin{equation}
\xi^+_{-\frac{1}{2}}(\vec{p}\,)\vec{\sigma}\cdot\vec{\epsilon}\,^*(\vec{n}_{\mathcal{P}},\lambda=0)\eta_{\frac{1}{2}}(\vec{p}\,')=\xi^+_{-\frac{1}{2}}(\vec{p}\,)\sigma_3\eta_{\frac{1}{2}}(\vec{p}\,')
=\sin\left(\frac{\alpha}{2}\right)\sin\left(\frac{\gamma}{2}\right)+\cos\left(\frac{\alpha}{2}\right)\cos\left(\frac{\gamma}{2}\right).
\end{equation}
The temporal components contribution to the probability is proportional with the bispinors product $\xi^+_{\sigma}(\vec{p}\,)\eta_{\sigma'}(\vec{p}\,')$ ,which is evaluated by using the momenta in the plane $(1,3)$ as above and the fact that $\sigma=-\frac{1}{2},\,\sigma'=\frac{1}{2}$ :
\begin{equation}
\xi^+_{-\frac{1}{2}}(\vec{p}\,)\eta_{\frac{1}{2}}(\vec{p}\,')=\sin\left(\frac{\alpha}{2}\right)\sin\left(\frac{\gamma}{2}\right)-\cos\left(\frac{\alpha}{2}\right)\cos\left(\frac{\gamma}{2}\right).
\end{equation}
Then collecting all the above results for $\lambda=0$, the sum containing the terms with longitudinal modes contribution give
\begin{eqnarray}\label{av}
|A(\lambda=0)|^2=|A_{k}(\mathcal{P},p,p')|^2\cos^2\left(\frac{\alpha+\gamma}{2}\right)+|C_{k}(\mathcal{P},p,p')|^2\cos^2\left(\frac{\alpha-\gamma}{2}\right)\nonumber\\
-(A_{k}^*(\mathcal{P},p,p')C_{k}(\mathcal{P},p,p')
+C_{k}^*(\mathcal{P},p,p')A_{k}(\mathcal{P},p,p'))\cos\left(\frac{\alpha+\gamma}{2}\right)\cos\left(\frac{\alpha-\gamma}{2}\right)
\end{eqnarray}
From the above equation we observe that if we take $\alpha=\gamma=0$ then the probability is nonvanishing and we conclude that the modes with $\lambda=0$ favour production processes in which the momenta of the neutrino and antineutrino are parallel and have the same orientation.

For the case with $\lambda=\pm1$, we consider the circular polarizations such that
\begin{equation}\label{cp}
\vec{\epsilon}_{\pm1}=\frac{1}{\sqrt{2}}(\pm\vec{e}_{1}+i\vec{e}_{2}),
\end{equation}
in the local orthogonal frame where $\vec{\mathcal{P}}=\mathcal{P}\vec{e}_3$. Then taking the spherical coordinates of the momenta $\vec{p}\,(p,\alpha,\beta=0)$ and $\vec{p}\,'(p',\gamma,\theta=\pi)$ the final result is:
\begin{eqnarray}\label{sh}
&&\xi^+_{-\frac{1}{2}}(\vec{p}\,)\vec{\sigma}\cdot\vec{\epsilon}\,^*(\vec{n}_{\mathcal{P}},\lambda=1)\eta_{\frac{1}{2}}(\vec{p}\,')=-2\cos\left(\frac{\alpha}{2}\right)\sin\left(\frac{\gamma}{2}\right),\,\lambda=1\nonumber\\
&&\xi^+_{-\frac{1}{2}}(\vec{p}\,)\vec{\sigma}\cdot\vec{\epsilon}\,^*(\vec{n}_{\mathcal{P}},\lambda=-1)\eta_{\frac{1}{2}}(\vec{p}\,')=-2\sin\left(\frac{\alpha}{2}\right)\cos\left(\frac{\gamma}{2}\right),\,\lambda=-1.
\end{eqnarray}
The above analysis completes our analytical results of the amplitude and probability equations and establish the formula for different values of the polar angles $\alpha,\,\gamma$. For example if one analyses the amplitude equation in the case when $\lambda=\pm1$ and we fix $\alpha=\gamma=0$ then a simple computation using equations (\ref{abc}) and (\ref{sh}) prove that the amplitude/probability are zero. This is the case when the fermions momenta are parallel and have the same orientation. We can conclude that only the modes with $\lambda=0$ will give contributions to the amplitudes for $\alpha=\gamma=0$.

Another aspect is related to the helicity conservation in the process of spontaneous production of Z boson and neutrino and antineutrino from vacuum. Since we obtain the selection rule for the helicities of the neutrino and atineutrino the only possible processes are restricted to $\sigma=-\frac{1}{2}$ and $\sigma'=\frac{1}{2}$. Then in the case with $\lambda=0$ the helicity is conserved, while for the case of transversal modes contribution $\lambda=\pm 1$ the helicity conservation law is broken.

Let us analyse the case when the momenta of the neutrino and antineutrino are parallel but they have opposite orientation
such that $\alpha=\pi,\,\gamma=0$ or $\alpha=0,\,\gamma=\pi$. Further we must take into account the momentum conservation. In this case one can choose for the Z boson the parameters $\vec{\mathcal{P}}=-\mathcal{P}\vec{e}_3,\,\lambda=\pm1$, the antineutrino parameters $\vec{p}\,'=p'\,\vec{e_3}$ with $\sigma'=\frac{1}{2}$ and neutrino parameter $\vec{p}=(\mathcal{P}-p')\vec{e_3}$, with $p'>\mathcal{P}$ and $\sigma=-\frac{1}{2}$. Then the final results for the amplitude obtained with the transversal modes $\lambda=\pm1$, is:
\begin{eqnarray}\label{l1}
\mathcal{A}(\lambda=\pm1)=\frac{-ie_0}{\sin(2\theta_{W})}\,\delta^3(\vec{\mathcal{P}}+\vec{p}+\vec{p}\,\,')\frac{1}{(2\pi)^{3/2}}\frac{4k^2+1}{2(2i\mathcal{P})^{3/2}\cosh(\pi k)}.
\end{eqnarray}
The situation analysed above proves that the processes in which the momenta of the fermions are parallel but have opposite orientation are favoured when $\lambda=\pm1$, while the contribution corresponding to the modes with $\lambda=0$ vanishes in this case. As a final observation we remark that the modes with $\lambda=\pm1$ give non-vanishing probabilities for processes which do not conserve the helicity.

\section{Total probability}
The total probability of the process can be obtained by integration after the final momenta of the resulting particles. The functions that define the amplitudes $A_{k}(\mathcal{P},p,p'),B_{k}(\mathcal{P},p,p'),C_{k}(\mathcal{P},p,p')$, are expressed in terms of hypergeometric functions whose algebraic argument depend on the modulus of momenta $p,p',\mathcal{P}$. The square modulus of the amplitude will contain products of two hypergeometric functions. A direct integration is impossible since these kind of integrals are not known in literature and in addition the computations are further complicated by the presence of angular integrals. Still in a realistic situation we can do computations by considering emission of "soft" Z bosons with small momenta modulus. Another trick that will help in our computation is to take the Z boson momenta vector on a fixed direction.
\subsection{Total probability for $\lambda=\pm1$}
Considering the contributions of the transversal modes ($\lambda=\pm1$) in total probability and using equation (\ref{pif1}) we obtain:

\begin{eqnarray}\label{f112}
&&P_{tot}(\lambda=\pm1)=\int d^3\mathcal{P}d^3p\,d^3p'\,P_{Z\nu\overline{\nu}}(\lambda=\pm1)=\int d^3\mathcal{P}d^3p\,d^3p'\,\frac{1}{2}\sum_{\lambda}|\mathcal{A}_{Z\nu\overline{\nu}}(\lambda=\pm1)|^2\nonumber\\
&&=\int d^3\mathcal{P}d^3p\,d^3p'\,\frac{e_0^2\,\delta^3(\vec{\mathcal{P}}+\vec{p}+\vec{p}\,\,')}{(2\pi)^{3}\sin^2(2\theta_{W})}\left(\frac{1}{2}-\sigma\right)^2\left(\frac{1}{2}+\sigma'\right)^2\nonumber\\
&&\times\frac{1}{2}\sum_{\lambda}|B_{k}(\mathcal{P},p,p')|^2|\xi^+_{\sigma}(\vec{p}\,)\vec{\sigma}\cdot\vec{\epsilon}\,^*(\vec{n}_{\mathcal{P}},\,\lambda=\pm1)\eta_{\sigma'}(\vec{p}\,')|^2
\nonumber\\
\end{eqnarray}
Then we take the momenta of Z boson fixed along the z axis, $\vec{\mathcal{P}}=\mathcal{P}\vec{e_3}$. Further using the conservation of the momentum the bispinor $\eta_{\frac{1}{2}}(\vec{p\,'})$ can be rewritten in terms of the momenta $p,\mathcal{P}$ ,following the integration after $d^3p\,'$. The next step is to set according to the above formulas the possible values of polarizations such that $\sigma=-\frac{1}{2}\,,\sigma'=\frac{1}{2}$.
We start with the general formula for $B_{k}(\mathcal{P},p,p')$ functions which contain all momenta dependence in amplitude, given in equation (\ref{cb}). Then perform the integration with the delta Dirac function after the momenta of the anti-neutrinos $d^3p'$:
\begin{eqnarray}
\int d^3p'\,\delta^3(\vec{\mathcal{P}}+\vec{p}+\vec{p}\,\,')|B_{k}(\mathcal{P},p,p')|^2|\xi^+_{\sigma}(\vec{p}\,)\vec{\sigma}\cdot\vec{\epsilon}\,^*(\vec{n}_{\mathcal{P}},\,\lambda=\pm1)\eta_{\sigma'}(\vec{p}\,')|^2\nonumber\\
=|B_{k}(\mathcal{P},p,|\vec{\mathcal{P}}+\vec{p}|\,)|^2
|\xi^+_{\sigma}(\vec{p}\,)\vec{\sigma}\cdot\vec{\epsilon}\,^*(\vec{n}_{\mathcal{P}},\,\lambda=\pm1)\eta_{\sigma'}(-(\vec{\mathcal{P}}+\vec{p}\,))|^2
\end{eqnarray}
The result of this integral depends on the new functions:
\begin{eqnarray}
&&|B_{k}(\mathcal{P},p,|\vec{\mathcal{P}}+\vec{p}|\,)|^2=\frac{1}{(\mathcal{P}+p+|\vec{\mathcal{P}}+\vec{p}|)^{3}}
\left|\Gamma\left(\frac{3}{2}-ik\right)\right|^2\left|\Gamma\left(\frac{3}{2}+ik\right)\right|^2\nonumber\\
&&\times\,_{2}F_{1}\left(\frac{3}{2}-i
k,\frac{1}{2}-ik;2;\frac{-\mathcal{P}+p+|\vec{\mathcal{P}}+\vec{p}|}{\mathcal{P}+p+|\vec{\mathcal{P}}+\vec{p}|}\right)
\,_{2}F_{1}\left(\frac{3}{2}+i
k,\frac{1}{2}+ik;2;\frac{-\mathcal{P}+p+|\vec{\mathcal{P}}+\vec{p}|}{\mathcal{P}+p+|\vec{\mathcal{P}}+\vec{p}|}\right).
\nonumber\\
\end{eqnarray}
\begin{eqnarray}\label{spinss}
&&\eta_{\frac{1}{2}}(-(\vec{p}+\vec{\mathcal{P}}\,))=\sqrt{\frac{\mathcal{P}+p\cos(\alpha)+|\vec{\mathcal{P}}+\vec{p}|}{2|\vec{\mathcal{P}}+\vec{p}|}}\left(
  \begin{array}{c}
   \frac{p\sin(\alpha)e^{-i\beta}}{\mathcal{P}+p\cos(\alpha)+|\vec{\mathcal{P}}+\vec{p}|} \\
    -1 \\
  \end{array}
\right).
\end{eqnarray}
\begin{eqnarray}\label{spinsss}
&&\xi_{-\frac{1}{2}}(\vec{p}\,)=\sqrt{\frac{1+\cos(\alpha)}{2}}\left(
  \begin{array}{c}
   \frac{-\sin(\alpha)e^{-i\beta}}{1+\cos(\alpha)} \\
    1 \\
  \end{array}
\right).
\end{eqnarray}
Equation (\ref{spinss}) was obtained taking into account that $\vec{\mathcal{P}}=\mathcal{P}\vec{e_3}$ and the conservation law of momentum $\vec{p}\,\,'=-(\vec{p}+\vec{\mathcal{P}})$, from which we obtain that the components on the $1,2$ axes are
$p_1=p_1'\,\,;p_2=p_2';\,\mathcal{P}_1=\mathcal{P}_2=0,\mathcal{P}_3=\mathcal{P} $. Now considering the spherical coordinates for the momenta $\vec{p}\,(p,\alpha,\beta)$ and $\vec{p}\,'(p',\gamma,\theta)$, together with the momenta conservation relations projected on axes we obtain: $p_3'=p'\cos(\gamma)=\mathcal{P}+p\cos(\alpha)$, $p'=|\vec{\mathcal{P}}+\vec{p}|$ and $p_1'-ip_2'=p_1-ip_2=p\sin(\alpha)e^{-i\beta}$, and finally a simple computation give equation (\ref{spinss}). In this way the helicity bispinor $\eta_{\frac{1}{2}}(\vec{p}\,\,')$, was expressed in terms of momenta $p,\mathcal{P}$, polar angle $\alpha$ and azimuthal angle $\beta$, thus allowing us to perform the further integration after $d^3p\,,d^3\mathcal{P}$. After a little calculation we obtain that:
\begin{eqnarray}\label{sc}
&&\xi^+_{-\frac{1}{2}}(\vec{p}\,)\vec{\sigma}\cdot\vec{\epsilon}\,^*(\vec{n}_{\mathcal{P}},\,\lambda=1)\eta_{\frac{1}{2}}(-(\vec{\mathcal{P}}+\vec{p}\,))=
\left(\frac{p\sin(\alpha)e^{-i\beta}\sqrt{1+\cos(\alpha)}}{\sqrt{2(\mathcal{P}+p\cos(\alpha)+|\vec{\mathcal{P}}+\vec{p}|)|\vec{\mathcal{P}}+\vec{p}|}}\right)\nonumber\\
&&\xi^+_{-\frac{1}{2}}(\vec{p}\,)\vec{\sigma}\cdot\vec{\epsilon}\,^*(\vec{n}_{\mathcal{P}},\,\lambda=-1)\eta_{\frac{1}{2}}(-(\vec{\mathcal{P}}+\vec{p}\,))=
\left(\frac{-\sin(\alpha)e^{i\beta}\sqrt{(\mathcal{P}+p\cos(\alpha)+|\vec{\mathcal{P}}+\vec{p}|)}}{\sqrt{2|\vec{\mathcal{P}}+\vec{p}|}\sqrt{1+\cos(\alpha)}}\right)
\nonumber\\
\end{eqnarray}
The dependence of the momenta sum modulus $|\vec{\mathcal{P}}+\vec{p}|$, will allow us to consider fixed directions by fixing the angle between momenta vectors $\vec{\mathcal{P}}$ and $\vec{p}$ which is $\alpha$ because the Z boson momenta is on the third axis.  Even so the result is expressed in terms of integrals with products of two hypergeometric functions and a polynomial factor in momenta powers, such that the momenta integrals that define the total probability are:
\begin{eqnarray}\label{fi}
&&\int d\Omega_{\mathcal{P}}\int_0^{\infty} d{\mathcal{P}}{\mathcal{P}}^2\int d\Omega_{p}\int_0^{\infty} dp\, \frac{p^2}{(\mathcal{P}+p+|\vec{\mathcal{P}}+\vec{p}|\,)^3}\,\nonumber\\
&&\times\,_{2}F_{1}\left(\frac{3}{2}-i
k,\frac{1}{2}-ik;2;\frac{-\mathcal{P}+p+|\vec{\mathcal{P}}+\vec{p}|}{\mathcal{P}+p+|\vec{\mathcal{P}}+\vec{p}|}\right)
\,_{2}F_{1}\left(\frac{3}{2}+i
k,\frac{1}{2}+ik;2;\frac{-\mathcal{P}+p+|\vec{\mathcal{P}}+\vec{p}|}{\mathcal{P}+p+|\vec{\mathcal{P}}+\vec{p}|}\right)\nonumber\\
&&\times\left\{
\begin{array}{cll}
\left(\frac{(1+\cos(\alpha))p^2\sin^2(\alpha)}{2(\mathcal{P}+p\cos(\alpha)+|\vec{\mathcal{P}}+\vec{p}|)|\vec{\mathcal{P}}+\vec{p}|}\right)
&{\rm for}&\lambda=1\\
\left(\frac{(\mathcal{P}+p\cos(\alpha)+|\vec{\mathcal{P}}+\vec{p}|)\sin^2(\alpha)}{2(1+\cos(\alpha))|\vec{\mathcal{P}}+\vec{p}|}\right)
&{\rm for}&\lambda=-1\\
\end{array}\right.
\end{eqnarray}
and these integrals do not have analytical expression to the best of our knowledge. Since the momenta of the Z boson is along the z axis then the solid angle integral $\int d\Omega_{\mathcal{P}}=4\pi$, while in the case of the angular integral corresponding to $p$ , we can solve the integral after the azimuthal angle $\beta$ and obtain :
\begin{eqnarray}
&&\int d\Omega_{p}=\int_{0}^{2\pi}d\beta\int_{-1}^{1} d(\cos\alpha)(...)=2\pi\int_{-1}^{1} d(\cos\alpha)(...)
\end{eqnarray}
Now because the integrand have a complicated dependence on the angle $\alpha$, we will consider the situation when the polar angle is fixed. The form of bispinor product given by equation (\ref{sc}) suggest that we can take $\alpha=\pi/2$, because for values of $\alpha=0,\pi$ the probability is vanishing for $\lambda=\pm1$. We consider the situation when $p>>\mathcal{P}$ and fixing polar angle to $\alpha=\pi/2$, the square modulus of bispinor products give :
\begin{eqnarray}\label{xin}
|\xi^+_{-\frac{1}{2}}(\vec{p}\,)\vec{\sigma}\cdot\vec{\epsilon}\,^*(\vec{n}_{\mathcal{P}},\,\lambda=1)\eta_{\frac{1}{2}}(-(\vec{\mathcal{P}}+\vec{p}\,))|^2=
\frac{p}{2(p+\mathcal{P})}\nonumber\\
|\xi^+_{-\frac{1}{2}}(\vec{p}\,)\vec{\sigma}\cdot\vec{\epsilon}\,^*(\vec{n}_{\mathcal{P}},\,\lambda=-1)\eta_{\frac{1}{2}}(-(\vec{\mathcal{P}}+\vec{p}\,))|^2=
\frac{(p+\mathcal{P})}{2p},
\end{eqnarray}
where the approximation $|\vec{\mathcal{P}}+\vec{p}|=\sqrt{p^2+\mathcal{P}^2}\simeq p\left(1+\frac{1}{2}\left(\frac{\mathcal{P}}{p}\right)^2\right)\sim p$, was used.
Let us turn now to the hypergeometric functions. If the definition of the hypergeometric functions is used, it is known that this behave as a series in powers of $\frac{-\mathcal{P}+p+|\vec{\mathcal{P}}+\vec{p}|}{\mathcal{P}+p+|\vec{\mathcal{P}}+\vec{p}|}$.
For obtaining the final result analytically we will make approximations using the fact that in a realistic situation the momenta of the produced massive Z particles will be much more smaller than the momenta of neutrinos. Now recalling the fact that the Z boson momenta can be  written with the well known relation in terms of wavelength, $\mathcal{P}=h/\lambda$. In this approximation the integration after the momenta modulus of the Z boson will be considered from zero up to the momenta corresponding to the Compton wavelength, and the upper limit of the integral will be cut to $\frac{\mathcal{P}_C}{\omega}$:
\begin{equation}
\frac{\mathcal{P}_C}{\omega}=\frac{h}{\omega\lambda_{C}}=\frac{M_Z\,c}{\omega},\, \lambda_{C}=\frac{h}{M_Z c}
\end{equation}
Where $\lambda_{C}$ is the Compton wavelength of the associated to Z boson, which is proportional with $M_Z^{-1}$, and from here we observe that the range of electro-weak force is very small, since the mass of the Z boson is large. In the above equation we can take $\hbar=c=1$ since the natural units were used.

The integrals after neutrino momentum $p$ are logarithmically divergent taking into account the integrand proportionality with $\frac{p^2}{(\mathcal{P}+p+|\vec{\mathcal{P}}+\vec{p}|\,)^3}$. Considering the above setup in which the angle between momenta vectors $\vec{\mathcal{P}}\,,\vec{p}$ is fixed at $\alpha=\pi/2$ and $|\vec{\mathcal{P}}+\vec{p}|\simeq p$ for $p>>\mathcal{P}$, the algebraic argument of hypergeometric functions become
\begin{equation}
\frac{-\mathcal{P}+p+|\vec{\mathcal{P}}+\vec{p}|}{\mathcal{P}+p+|\vec{\mathcal{P}}+\vec{p}|}\simeq\frac{2p-\mathcal{P}}{2p+\mathcal{P}}\sim1,
\end{equation}
and we can use the expansion of hypergeometric function $_{2}F_{1}(a,b;c;1)=\frac{\Gamma(c)\Gamma(c-a-b)}{\Gamma(c-a)\Gamma(c-b)}$ to obtain:
\begin{equation}\label{hipe}
_{2}F_{1}\left(\frac{3}{2}-i
k,\frac{1}{2}-ik;2;1\right)= \frac{\Gamma(2)\Gamma(2ik)}{\Gamma\left(\frac{3}{2}+ik\right)\Gamma\left(\frac{1}{2}+ik\right)}.
\end{equation}
In this setup equation (\ref{fi}) for the momenta integrals can be rewritten by using equations (\ref{xin}) and (\ref{hipe}) and we mention that the result of the angular integrals is included:
\begin{eqnarray}\label{fin}
\frac{\pi^2}{2}\left|\frac{\Gamma(2ik)}{\Gamma\left(\frac{3}{2}+ik\right)\Gamma\left(\frac{1}{2}+ik\right)}\right|^2
\left\{
\begin{array}{cll}
\int_0^{\mathcal{P}_C/\omega} d{\mathcal{P}}{\mathcal{P}}^2\int_0^{\infty} dp\, \frac{p^3}{(\mathcal{P}+p)^4}
&{\rm for}&\lambda=1\\
\int_0^{\mathcal{P}_C/\omega} d{\mathcal{P}}{\mathcal{P}}^2\int_0^{\infty} dp\, \frac{p}{(\mathcal{P}+p)^2}
&{\rm for}&\lambda=-1\\
\end{array}\right.
\end{eqnarray}
The first step here will be to solve the infinite integral after $p$ which is logarithmicaly divergent. Next we observe that the integrands in equation (\ref{fin}) could be expressed with the new indefinite integrals after a new variable $y$:
\begin{equation}\label{tr}
-4\int\frac{dy\,p^3}{(\mathcal{P}+p+y)^5}=\frac{p^3}{(\mathcal{P}+p+y)^4};\,\,-2\int\frac{dy\,p}{(\mathcal{P}+p+y)^3}=\frac{p}{(\mathcal{P}+p+y)^2}
\end{equation}
which in the limit $y\rightarrow0$, reduce to the original integrand functions $\frac{p^3}{(\mathcal{P}+p)^4}$ and $\frac{p}{(\mathcal{P}+p)^2}$. The second step is to solve all the integrals including the integral after the new variable $y$, and in the end to take the limit $y\rightarrow0$. Now we replace equation (\ref{tr}) in equation (\ref{fin}), then solve first the integral after neutrino momentum $p$ which is no longer divergent, and observe that the last two integrals after variables $\mathcal{P},y$ contain only elementary functions. The final result is obtained by taking the limit $y\rightarrow0$ as it was proven in equation (\ref{divi}) from Appendix:
\begin{eqnarray}\label{fine}
&&\lim_{y\rightarrow0}\left[-4\int dy\int_0^{\mathcal{P}_{C}/\omega}d{\mathcal{P}}{\mathcal{P}}^2  \int_0^{\infty} dp\,\frac{p^3}{(\mathcal{P}+p+y)^5}\right]\nonumber\\
&&=\lim_{y\rightarrow0}\left[-2\int dy\int_0^{\mathcal{P}_{C}/\omega}d{\mathcal{P}}{\mathcal{P}}^2  \int_0^{\infty} dp\,\frac{p}{(\mathcal{P}+p+y)^3}\right]=\left(\frac{\mathcal{P}_{C}}{\omega}\right)^3\left(\frac{11}{18}-\frac{1}{3}\ln\left(\frac{\mathcal{P}_{C}}{\omega}\right)\right).
\end{eqnarray}
Our result prove that the total probabilities corresponding to $\lambda=1$ and $\lambda=-1$ are equal.
The final expression for the total probability is obtained after replacing the results of the momentum integrals in equation (\ref{f112}) and observe that we can introduce the fine constant structure $\alpha=\frac{e_0^2}{(4\pi)}$ and the dependence on $M_Z/\omega$, by taking into account that $\frac{\mathcal{P}_{C}}{\omega}=\frac{M_Z}{\omega}$:
\begin{eqnarray}\label{fina}
&&P_{tot}(\lambda=\pm1)=\frac{\alpha}{8\sin^2(2\theta_{W})}\,\frac{\left|\Gamma\left(\frac{3}{2}-ik\right)\right|^2}{\left|\Gamma\left(\frac{1}{2}-ik\right)\right|^2}\left|\Gamma\left(2ik\right)\right|^2
\left(\frac{M_{Z}}{\omega}\right)^3\left(\frac{11}{18}-\frac{1}{3}\ln\left(\frac{M_{Z}}{\omega}\right)\right).
\end{eqnarray}
The outcome of our computations is the total probability, and the equation (\ref{fina}) gives this quantity for the process in which the triplet Z boson, neutrino and anti-neutrino are generated from de Sitter vacuum. Remarkably is that the total probability depends only on the ratio $M_Z/\omega$. Let us study the behaviour of the total probability in terms of the parameter $M_Z/\omega$ by a graphical analysis.
\begin{figure}[h!t]
\includegraphics[scale=0.4]{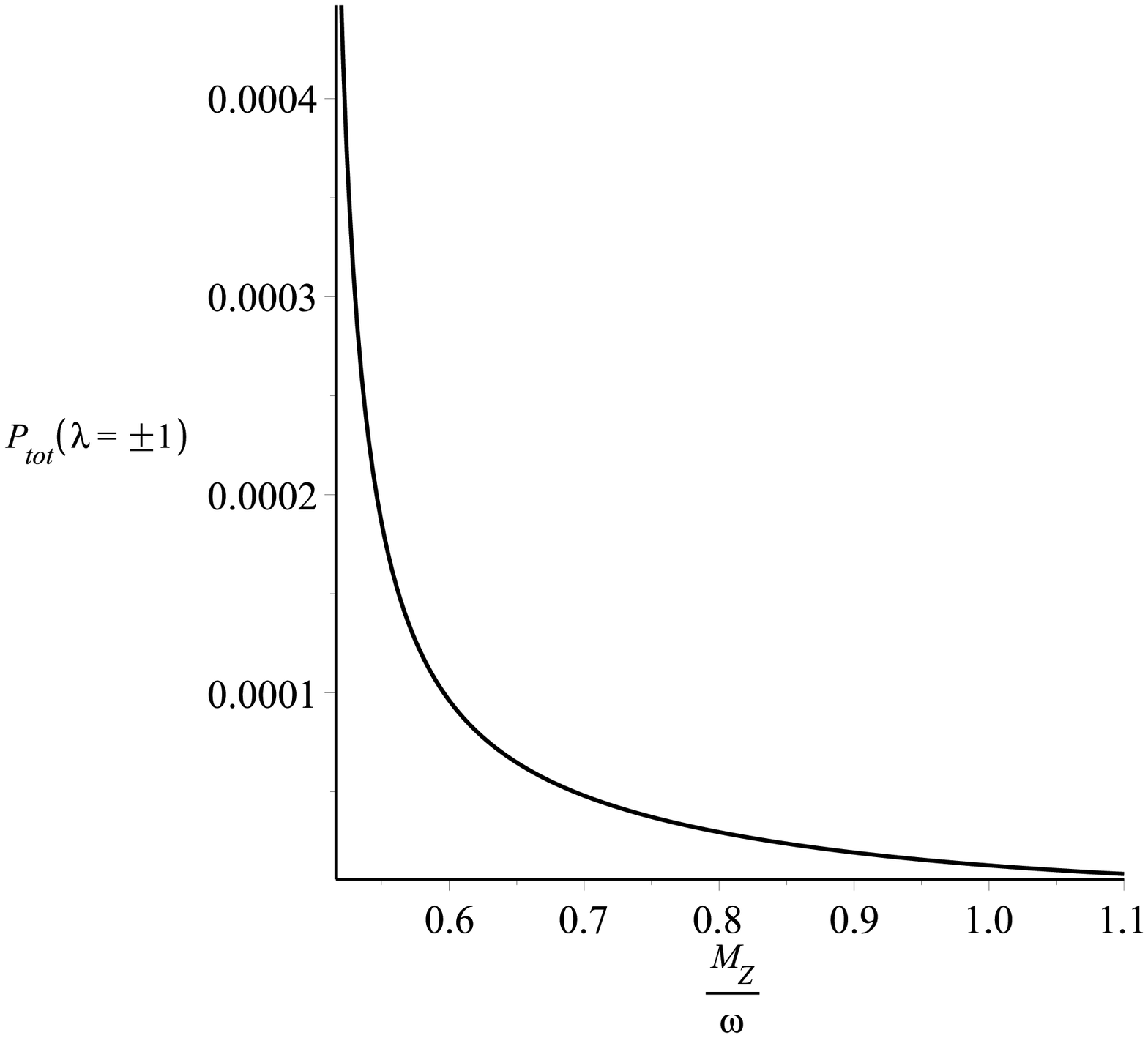}
\caption{$P_{tot}(\lambda=\pm1)$ as a function of parameter $M_Z/\omega$.}
\label{f9}
\end{figure}
The figure (\ref{f9}) shows that the total probability computed by using the transversal modes will remain a function that is nonvanishing only when the parameter $M_Z/\omega$ is small and in the Minkowski limit the total probability is vanishing. This behaviour is due to the fact that the Euler gamma functions $\left|\Gamma\left(2ik\right)\right|^2 $ can be written in terms of $sinh^{-1}(2\pi k)$ and this factor will contain the nonvanishing values of the total probability in a small interval around $M_Z/\omega\leq 1$. In the limit $M_Z/\omega >>1$ the factor $sinh^{-1}(2\pi k)\sim e^{-2\pi M_Z/\omega}$ and the total probability will drop quickly to zero. This behaviour was also observed in the graphs of the probability density given in the previous section.

An alternative method to obtain the total probability could be like follows. First use the relation (\ref{hy}) from appendix and rewrite the hypergeometric functions.
Then the relation $_{2}F_{1}(a,b;2b;z)\approx(1-z)^{-a}$, can be used. This is a good approximation, which is close to the exact formula $_{2}F_{1}(a,b;b;z)=(1-z)^{-a}$, and this can be checked using the definition of the hypergeometric function and plot the real part and imaginary part of hypergeometric function and compare with the plot of polynomial approximation $(1-z)^{-a}$, where $a$ will depend on parameter $k$. Still one need to make detailed computations for obtaining a valid equation. We hope to study this method in a future paper.

\subsection{Total probability for $\lambda=0$}
The contribution of the longitudinal modes with $\lambda=0$, to the total probability can be obtained integrating after the final momenta the probability given in equation (\ref{pif}):

\begin{eqnarray}\label{pif2}
P_{tot}(\lambda=0)&=&\int d^3\mathcal{P}d^3p\,d^3p'\,P_{Z\nu\overline{\nu}}(\lambda=0)=\int d^3\mathcal{P}d^3p\,d^3p'\,
\frac{e_0^2}{\sin^2(2\theta_{W})}\,\delta^3(\vec{\mathcal{P}}+\vec{p}+\vec{p}\,\,')\nonumber\\
&&\frac{1}{(2\pi)^{3}}\left(\frac{1}{2}-\sigma\right)^2\left(\frac{1}{2}+\sigma'\right)^2\biggl\{
|A_{k}(\mathcal{P},p,p')|^2|\xi^+_{\sigma}(\vec{p}\,)\vec{\sigma}\cdot\vec{\epsilon}\,^*(\vec{n}_{\mathcal{P}},\lambda=0)\eta_{\sigma'}(\vec{p}\,')|^2\nonumber\\
&&+|C_{k}(\mathcal{P},p,p')|^2|\xi^+_{\sigma}(\vec{p}\,)\eta_{\sigma'}(\vec{p}\,')|^2\nonumber\\
&&+A_{k}^*(\mathcal{P},p,p')C_{k}(\mathcal{P},p,p')(\xi^+_{\sigma}(\vec{p}\,)\vec{\sigma}\cdot\vec{\epsilon}\,^*(\vec{n}_{\mathcal{P}},\lambda=0)\eta_{\sigma'}(\vec{p}\,'))^*(\xi^+_{\sigma}(\vec{p}\,)\eta_{\sigma'}(\vec{p}\,'))\nonumber\\
&&+C_{k}^*(\mathcal{P},p,p')A_{k}(\mathcal{P},p,p')(\xi^+_{\sigma}(\vec{p}\,)\eta_{\sigma'}(\vec{p}\,'))^*(\xi^+_{\sigma}(\vec{p}\,)\vec{\sigma}\cdot\vec{\epsilon}\,^*(\vec{n}_{\mathcal{P}},\lambda=0)\eta_{\sigma'}(\vec{p}\,'))
\biggl\}.
\end{eqnarray}
Then the analysis of the total probability is basically the same as in the case of transversal modes, and by solving the integral after $p'$ we obtain that the functions $|A_k|^2,|C_k|^2,A_k^*C_k,A_kC_k^*$ defining the probability will depend on modulus $|\vec{\mathcal{P}}+\vec{p}|$. The integrals after $d^3p'$ given in equation (\ref{pif2}), are solved by using use the delta Dirac function properties as follows:
\begin{eqnarray}\label{xii}
&&\int d^3p'\,\delta^3(\vec{\mathcal{P}}+\vec{p}+\vec{p}\,\,')\biggl\{
|A_{k}(\mathcal{P},p,p')|^2|\xi^+_{\sigma}(\vec{p}\,)\vec{\sigma}\cdot\vec{\epsilon}\,^*(\vec{n}_{\mathcal{P}},\lambda=0)\eta_{\sigma'}(\vec{p}\,')|^2\nonumber\\
&&+|C_{k}(\mathcal{P},p,p')|^2|\xi^+_{\sigma}(\vec{p}\,)\eta_{\sigma'}(\vec{p}\,')|^2\nonumber\\
&&+A_{k}^*(\mathcal{P},p,p')C_{k}(\mathcal{P},p,p')(\xi^+_{\sigma}(\vec{p}\,)\vec{\sigma}\cdot\vec{\epsilon}\,^*(\vec{n}_{\mathcal{P}},\lambda=0)\eta_{\sigma'}(\vec{p}\,'))^*(\xi^+_{\sigma}(\vec{p}\,)\eta_{\sigma'}(\vec{p}\,'))\nonumber\\
&&+C_{k}^*(\mathcal{P},p,p')A_{k}(\mathcal{P},p,p')(\xi^+_{\sigma}(\vec{p}\,)\eta_{\sigma'}(\vec{p}\,'))^*(\xi^+_{\sigma}(\vec{p}\,)\vec{\sigma}\cdot\vec{\epsilon}\,^*(\vec{n}_{\mathcal{P}},\lambda=0)\eta_{\sigma'}(\vec{p}\,'))
\biggl\}\nonumber\\
&&=|A_{k}(\mathcal{P},p,|\vec{\mathcal{P}}+\vec{p}|\,)|^2|\xi^+_{\sigma}(\vec{p}\,)\vec{\sigma}\cdot\vec{\epsilon}\,^*(\vec{n}_{\mathcal{P}},\lambda=0)\eta_{\sigma'}(-(\vec{\mathcal{P}}+\vec{p}\,))|^2\nonumber\\
&&+|C_{k}(\mathcal{P},p,|\vec{\mathcal{P}}+\vec{p}|\,)|^2|\xi^+_{\sigma}(\vec{p}\,)\eta_{\sigma'}(-(\vec{\mathcal{P}}+\vec{p}\,))|^2\nonumber\\
&&+A_{k}^*(\mathcal{P},p,|\vec{\mathcal{P}}+\vec{p}|\,)C_{k}(\mathcal{P},p,|\vec{\mathcal{P}}+\vec{p}|\,)(\xi^+_{\sigma}(\vec{p}\,)\vec{\sigma}\cdot\vec{\epsilon}\,^*(\vec{n}_{\mathcal{P}},\lambda=0)\eta_{\sigma'}(-(\vec{\mathcal{P}}+\vec{p}\,)))^*(\xi^+_{\sigma}(\vec{p}\,)\eta_{\sigma'}(-(\vec{\mathcal{P}}+\vec{p}\,)))\nonumber\\
&&+C_{k}^*(\mathcal{P},p,|\vec{\mathcal{P}}+\vec{p}|\,)A_{k}(\mathcal{P},p,|\vec{\mathcal{P}}+\vec{p}|\,)(\xi^+_{\sigma}(\vec{p}\,)\eta_{\sigma'}(-(\vec{\mathcal{P}}+\vec{p}\,)))^*(\xi^+_{\sigma}(\vec{p}\,)\vec{\sigma}\cdot\vec{\epsilon}\,^*(\vec{n}_{\mathcal{P}},\lambda=0)\eta_{\sigma'}(-(\vec{\mathcal{P}}+\vec{p}\,)))
\nonumber\\
\end{eqnarray}
The helicity bispinors products from equation (\ref{xii}) need to be computed. For that we know that the polarization vector is on the momentum direction $\vec{\epsilon}\,(\vec{n}_{\mathcal{P}},\lambda=0)=\vec{e}_3$ and by using equations (\ref{spinss}), (\ref{spinsss}) the final result reads:

\begin{eqnarray}\label{sq}
&&\xi^+_{-\frac{1}{2}}(\vec{p}\,)\vec{\sigma}\cdot\vec{\epsilon}\,^*(\vec{n}_{\mathcal{P}},\lambda=0)\eta_{\frac{1}{2}}(-(\vec{\mathcal{P}}+\vec{p}\,))=
\sqrt{\frac{\mathcal{P}+p\cos(\alpha)+|\vec{\mathcal{P}}+\vec{p}|}{2|\vec{\mathcal{P}}+\vec{p}|}}\sqrt{\frac{1+\cos(\alpha)}{2}}\nonumber\\
&&\times\left(\frac{-p\sin^2(\alpha)}{(1+\cos(\alpha))(\mathcal{P}+p\cos(\alpha)+|\vec{\mathcal{P}}+\vec{p}|)}+1\right);\nonumber\\
&&\xi^+_{-\frac{1}{2}}(\vec{p}\,)\eta_{\frac{1}{2}}(-(\vec{\mathcal{P}}+\vec{p}\,))=\sqrt{\frac{\mathcal{P}+p\cos(\alpha)+|\vec{\mathcal{P}}+\vec{p}|}{2|\vec{\mathcal{P}}+\vec{p}|}}\sqrt{\frac{1+\cos(\alpha)}{2}}\nonumber\\
&&\times\left(\frac{-p\sin^2(\alpha)}{(1+\cos(\alpha))(\mathcal{P}+p\cos(\alpha)+|\vec{\mathcal{P}}+\vec{p}|)}-1\right).
\end{eqnarray}
The other combinations with helicity bispinors from probability (\ref{pif2}), can be obtained form the above relations. The dependence of polar angle $\alpha$, in equation (\ref{sq}) is complicated and for this reason we will follow the method as in the previous case and fix the angle between $\vec{p},\vec{\mathcal{P}}$ to $\alpha=0$ such that  $|\vec{\mathcal{P}}+\vec{p}|=\mathcal{P}+p$, and the Z boson momenta oriented along the z axis. Then the bispinors square modulus from our probability gives, if we use equation(\ref{sq}) for $\alpha=0$:
\begin{eqnarray}
&&|\xi^+_{-\frac{1}{2}}(\vec{p}\,)\vec{\sigma}\cdot\vec{\epsilon}\,^*(\vec{n}_{\mathcal{P}},\lambda=0)\eta_{\frac{1}{2}}(-(\vec{\mathcal{P}}+\vec{p}\,))|^2
=|\xi^+_{-\frac{1}{2}}(\vec{p}\,)\eta_{\frac{1}{2}}(-(\vec{\mathcal{P}}+\vec{p}\,))|^2=1;\nonumber\\
&&(\xi^+_{-\frac{1}{2}}(\vec{p}\,)\eta_{\frac{1}{2}}(-(\vec{\mathcal{P}}+\vec{p}\,)))^*(\xi^+_{-\frac{1}{2}}(\vec{p}\,)\vec{\sigma}\cdot\vec{\epsilon}\,^*(\vec{n}_{\mathcal{P}},\lambda=0)\eta_{\frac{1}{2}}(-(\vec{\mathcal{P}}+\vec{p}\,)))\nonumber\\
&&=(\xi^+_{-\frac{1}{2}}(\vec{p}\,)\eta_{\frac{1}{2}}(-(\vec{\mathcal{P}}+\vec{p}\,)))(\xi^+_{-\frac{1}{2}}(\vec{p}\,)\vec{\sigma}\cdot\vec{\epsilon}\,^*(\vec{n}_{\mathcal{P}},\lambda=0)\eta_{\frac{1}{2}}(-(\vec{\mathcal{P}}+\vec{p}\,)))^*=-1.
\end{eqnarray}

The algebraic arguments of the hypergeometric functions for $\alpha=0$ become $\frac{-\mathcal{P}+p+|\vec{\mathcal{P}}+\vec{p}|}{\mathcal{P}+p+|\vec{\mathcal{P}}+\vec{p}|}=\frac{p}{\mathcal{P}+p}$.
Then the total probability in the case $\lambda =0$, if we replace the contributions of functions $A_k,C_k$ from equation (\ref{pif2}) will contain the following momenta integrals:

\begin{eqnarray}\label{fip}
&&\int d\Omega_{\mathcal{P}}\int_0^{\mathcal{P}_C/\omega} d{\mathcal{P}}{\mathcal{P}}^2\int d\Omega_{p}\int_0^{\infty} dp\, p^2\nonumber\\
&&\biggl\{\frac{A_1}{2^3(\mathcal{P}+p\,)^3}\,_{2}F_{1}\left(\frac{3}{2}-i
k,\frac{1}{2}-ik;2;\frac{p}{\mathcal{P}+p}\right)
\,_{2}F_{1}\left(\frac{3}{2}+i
k,\frac{1}{2}+ik;2;\frac{p}{\mathcal{P}+p}\right)\nonumber\\
&&+\frac{A_2\,\mathcal{P}^4}{2^7(\mathcal{P}+p\,)^7}\,_{2}F_{1}\left(\frac{7}{2}-i
k,\frac{3}{2}-ik;3;\frac{p}{\mathcal{P}+p}\right)
\,_{2}F_{1}\left(\frac{7}{2}+i
k,\frac{3}{2}+ik;3;\frac{p}{\mathcal{P}+p}\right)\nonumber\\
&&-\frac{A_3\,\mathcal{P}^2}{2^5(\mathcal{P}+p\,)^5}\,_{2}F_{1}\left(\frac{7}{2}+i
k,\frac{3}{2}+ik;3;\frac{p}{\mathcal{P}+p}\right)
\,_{2}F_{1}\left(\frac{3}{2}-i
k,\frac{1}{2}-ik;2;\frac{p}{\mathcal{P}+p}\right)\nonumber\\
&&-\frac{A_4\,\mathcal{P}^2}{2^5(\mathcal{P}+p\,)^5}\,_{2}F_{1}\left(\frac{7}{2}-i
k,\frac{3}{2}-ik;3;\frac{p}{\mathcal{P}+p}\right)
\,_{2}F_{1}\left(\frac{3}{2}+i
k,\frac{1}{2}+ik;2;\frac{p}{\mathcal{P}+p}\right)\nonumber\\
&&+\frac{A_5\,\mathcal{P}^2}{2^7(\mathcal{P}+p\,)^5}\,_{2}F_{1}\left(\frac{5}{2}+i
k,\frac{1}{2}+ik;3;\frac{p}{\mathcal{P}+p}\right)
\,_{2}F_{1}\left(\frac{5}{2}-i
k,\frac{1}{2}-ik;3;\frac{p}{\mathcal{P}+p}\right)\nonumber\\
&&-\frac{A_6\,\mathcal{P}}{2^5(\mathcal{P}+p\,)^4}\,_{2}F_{1}\left(\frac{5}{2}+i
k,\frac{1}{2}+ik;3;\frac{p}{\mathcal{P}+p}\right)
\,_{2}F_{1}\left(\frac{3}{2}-i
k,\frac{1}{2}-ik;2;\frac{p}{\mathcal{P}+p}\right)\nonumber\\
&&+\frac{A_7\,\mathcal{P}^3}{2^7(\mathcal{P}+p\,)^6}\,_{2}F_{1}\left(\frac{5}{2}+i
k,\frac{1}{2}+ik;3;\frac{p}{\mathcal{P}+p}\right)
\,_{2}F_{1}\left(\frac{7}{2}-i
k,\frac{3}{2}-ik;3;\frac{p}{\mathcal{P}+p}\right)\nonumber\\
&&-\frac{A_8\,\mathcal{P}}{2^5(\mathcal{P}+p\,)^4}\,_{2}F_{1}\left(\frac{3}{2}+i
k,\frac{1}{2}+ik;2;\frac{p}{\mathcal{P}+p}\right)
\,_{2}F_{1}\left(\frac{5}{2}-i
k,\frac{1}{2}-ik;3;\frac{p}{\mathcal{P}+p}\right)\nonumber\\
&&+\frac{A_9\,\mathcal{P}^3}{2^7(\mathcal{P}+p\,)^6}\,_{2}F_{1}\left(\frac{7}{2}+i
k,\frac{3}{2}+ik;3;\frac{p}{\mathcal{P}+p}\right)
\,_{2}F_{1}\left(\frac{5}{2}-i
k,\frac{1}{2}-ik;3;\frac{p}{\mathcal{P}+p}\right)\biggl\},
\end{eqnarray}
where the coefficients of each integral $A_1...A_9$ are given by:
\begin{eqnarray}\label{coef1}
&&A_1=\left|\Gamma\left(\frac{3}{2}-ik\right)\right|^2\left|\Gamma\left(\frac{3}{2}+ik\right)\right|^2\left(\frac{1}{4}+k^2\right)\,,
A_2=\left|\Gamma\left(\frac{7}{2}-ik\right)\right|^2\left|\Gamma\left(\frac{3}{2}+ik\right)\right|^2\nonumber\\
&&A_3=\Gamma^2\left(\frac{3}{2}-ik\right)\Gamma\left(\frac{3}{2}+ik\right)\Gamma\left(\frac{7}{2}+ik\right)\left(\frac{1}{2}-ik\right),\nonumber\\
&&A_4=\Gamma^2\left(\frac{3}{2}+ik\right)\Gamma\left(\frac{3}{2}-ik\right)\Gamma\left(\frac{7}{2}-ik\right)\left(\frac{1}{2}+ik\right),\nonumber\\
&&A_5=\left|\Gamma\left(\frac{5}{2}-ik\right)\right|^2\left|\Gamma\left(\frac{5}{2}+ik\right)\right|^2\,,
A_6=\left|\Gamma\left(\frac{5}{2}-ik\right)\right|^2\left|\Gamma\left(\frac{3}{2}-ik\right)\right|^2\left(\frac{1}{2}-ik\right),\nonumber\\
&&A_7=\left|\Gamma\left(\frac{5}{2}-ik\right)\right|^2\Gamma\left(\frac{3}{2}+ik\right)\Gamma\left(\frac{7}{2}-ik\right)\,,
A_8=\left|\Gamma\left(\frac{5}{2}-ik\right)\right|^2\left|\Gamma\left(\frac{3}{2}-ik\right)\right|^2\left(\frac{1}{2}+ik\right),\nonumber\\
&&A_9=\left|\Gamma\left(\frac{5}{2}-ik\right)\right|^2\Gamma\left(\frac{3}{2}-ik\right)\Gamma\left(\frac{7}{2}+ik\right).
\end{eqnarray}
The integrals from equation (\ref{fip}) can be solved by taking the same approximation $p>>\mathcal{P}$ to simplify the algebraic argument of the hypergeometric function to $\frac{p}{\mathcal{P}+p}\sim1$, and each of the hypergeometric functions from the momenta integrals can be rewritten by using $_{2}F_{1}(a,b;c;1)=\frac{\Gamma(c)\Gamma(c-a-b)}{\Gamma(c-a)\Gamma(c-b)}$. In this way we arrive at integrals of the form given in eqs. (\ref{divi}),\,(\ref{plo}) from Appendix. Then final equation for the total probability in the case $\lambda=0$ is :
\begin{eqnarray}\label{p0}
&&P_{tot}(\lambda=0)=\frac{4\alpha}{\sin^2(2\theta_{W})}\left(\frac{M_Z}{\omega}\right)\biggl\{\frac{A_1B_1}{2^3}
\left(\frac{11}{18}-\frac{1}{3}\ln\left(\frac{M_{Z}}{\omega}\right)\right)+\frac{A_2B_2}{180\cdot2^7}\nonumber\\
&&-\frac{A_3B_3}{36\cdot2^5}-\frac{A_4B_4}{36\cdot2^5}+\frac{A_5B_5}{36\cdot2^7}
-\frac{A_6B_6}{9\cdot2^5}+\frac{A_7B_7}{90\cdot2^7}-\frac{A_8B_8}{9\cdot2^5}+\frac{A_9B_9}{90\cdot2^7}\biggl\}.
\end{eqnarray}
In equation (\ref{p0}) the coefficients $A_1...A_9$  are defined in equation (\ref{coef1}). The new coefficients $B_1...B_9$ resulted from the hypergeometric functions have the following expression:
\begin{eqnarray}\label{coef2}
&&B_1=\frac{\left|\Gamma\left(2ik\right)\right|^2}{\left|\Gamma\left(\frac{1}{2}+ik\right)\right|^2\left|\Gamma\left(\frac{3}{2}+ik\right)\right|^2}\,,
B_2=\frac{4\left|\Gamma\left(-2+2ik\right)\right|^2}{\left|\Gamma\left(-\frac{1}{2}+ik\right)\right|^2\left|\Gamma\left(\frac{3}{2}+ik\right)\right|^2}\nonumber\\
&&B_3=\frac{2\Gamma\left(-2-2ik\right)\Gamma\left(2ik\right)}{|\Gamma\left(\frac{3}{2}-ik\right)|^2\Gamma\left(\frac{1}{2}+ik\right)\Gamma\left(-\frac{1}{2}-ik\right)},
B_4=\frac{2\Gamma\left(-2+2ik\right)\Gamma\left(-2ik\right)}{|\Gamma\left(\frac{3}{2}-ik\right)|^2\Gamma\left(\frac{1}{2}-ik\right)\Gamma\left(-\frac{1}{2}+ik\right)},\nonumber\\
&&B_5=\frac{4\left|\Gamma\left(2ik\right)\right|^2}{\left|\Gamma\left(\frac{1}{2}+ik\right)\right|^2\left|\Gamma\left(\frac{5}{2}+ik\right)\right|^2}\,,
B_6=\frac{2\left|\Gamma\left(2ik\right)\right|^2}{\left|\Gamma\left(\frac{1}{2}+ik\right)\right|^2\Gamma\left(\frac{3}{2}+ik\right)\Gamma\left(\frac{5}{2}-ik\right)},\nonumber\\
&&B_7=\frac{4\Gamma\left(-2+2ik\right)\Gamma\left(-2ik\right)}{\Gamma\left(\frac{3}{2}+ik\right)\Gamma\left(\frac{5}{2}-ik\right)\Gamma\left(\frac{1}{2}-ik\right)\Gamma\left(-\frac{1}{2}+ik\right)}\,,
B_8=\frac{2\left|\Gamma\left(2ik\right)\right|^2}{\left|\Gamma\left(\frac{1}{2}+ik\right)\right|^2\Gamma\left(\frac{3}{2}-ik\right)\Gamma\left(\frac{5}{2}+ik\right)},\nonumber\\
&&B_9=\frac{4\Gamma\left(-2-2ik\right)\Gamma\left(2ik\right)}{\Gamma\left(\frac{3}{2}-ik\right)\Gamma\left(\frac{5}{2}+ik\right)\Gamma\left(\frac{1}{2}+ik\right)\Gamma\left(-\frac{1}{2}-ik\right)}.
\end{eqnarray}
The above equation give the contribution of the longitudinal modes to the total probability and this quantity is again dependent on the ratio $M_Z/\omega$. Next one can study the behaviour of the total probability in terms of the parameter $M_Z/\omega$,
\begin{figure}[h!t]
\includegraphics[scale=0.4]{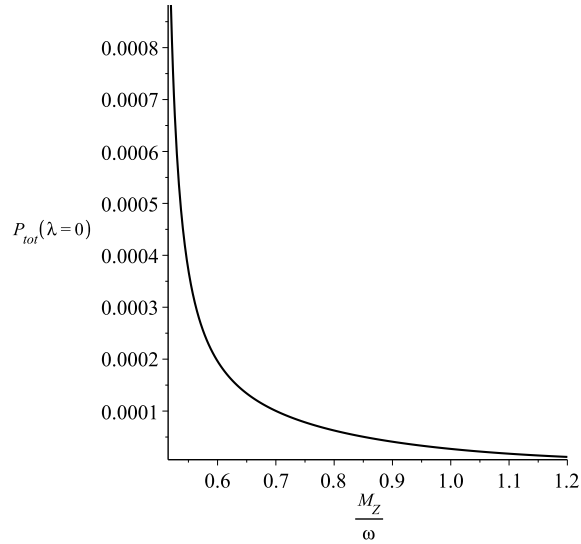}
\caption{$P_{tot}(\lambda=0)$ as a function of parameter $M_Z/\omega$.}
\label{f10}
\end{figure}
From the graph (\ref{f10}), we observe that the total probability for $\lambda=0$ is vanishing in the Minkowski limit. Another observation is that generation of helicity zero bosons is possible only when the gravitational fields are strong. Looking for the behaviour of the total probability in the case $\lambda=0$, for $M_Z/\omega>>1$, we must pay attention to the coefficients $A_1,B_1...A_9,B_9$ which define the total probability. A simple calculation proves that all these coefficients depend on Euler gamma functions which can be expressed in terms of hyperbolic functions $sinh^{-2}(\pi k),cosh^{-2}(\pi k)$ making the probability significative only in the interval $M_Z/\omega\leq 1$. Another aspect of our graphical analysis is related to the fact that the total probability in the case $\lambda=\pm1$ is to the same order with the total probability for $\lambda=0$ ( see figs.(\ref{f9})-(\ref{f10})).

At the end of this section we will analyse by using a graphical method the situation when the condition $\frac{M_Z}{\omega}>\frac{1}{2}$ is no longer mandatory and the index of the Hankel functions become real. The analysis is restricted only to the modes with $\lambda=\pm1$, because in this case the solutions have a purely imaginary index $i\sqrt{\left(\frac{M_Z}{\omega}\right)^2-\frac{1}{4}}$. Then in the interval $\frac{M_Z}{\omega}>\frac{1}{2}\in[0,1/2]$ the index of Hankel functions $\sqrt{\frac{1}{4}-\left(\frac{M_Z}{\omega}\right)^2}$, will become real. A simple way to observe the behaviour of the interest quantities in this interval is to plot the square modulus of the functions $B_k$ that define the probability density, in terms of $\frac{M_Z}{\omega}>\frac{1}{2}$.
\begin{figure}[h!t]
\includegraphics[scale=0.4]{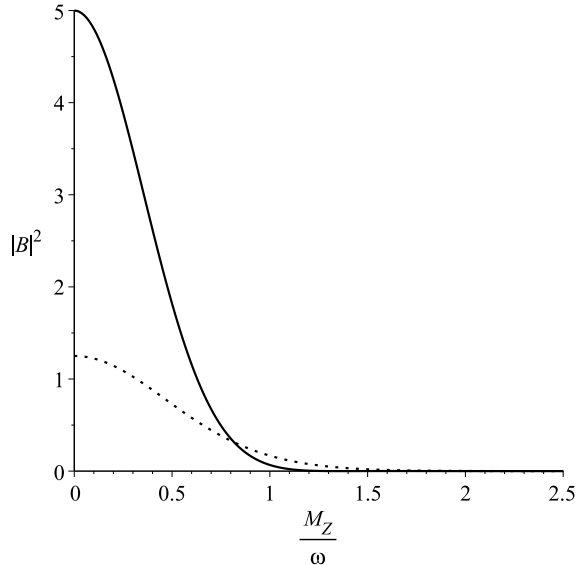}
\caption{ $|B_{k}|^2$ as a function of parameter $M_Z/\omega$. Solid line is for $p=0.5,p'=0.4,\mathcal{P}=0.1$, while the point line is for $p=0.5,p'=0.3,\mathcal{P}=0.2$ .}
\label{f11}
\end{figure}
We observe from our graph (\ref{f11}), that the probability density  will increase significatively in the interval $[0,1/2]$, by comparing with the case when solutions of the Proca equation were written with Hankel function of imaginary index. However a fully treatment of the problem will require an integration after the final momenta to obtain the total probability, and we hope to study this problem in detail in a future paper.
Here we just prove that our results can be extended to all values of parameter $\frac{M_Z}{\omega}$ and the same observations remains valid if the analysis is done with the modes corresponding to $\lambda=0$.

\section{Conclusions}
In this paper we study the interactions between Z boson and leptons in a de Sitter geometry by following the methods from flat space theory, which are based on construction of the transition amplitudes by using perturbations. The process in which the triplet Z boson and  neutrino-antineutrino pair are generated from vacuum in de Sitter geometry was studied by computing the first order transition amplitude corresponding to the neutral current coupling with the Z boson field. The basic steps of our computations were the exact solution of the Proca and Dirac equation in de Sitter geometry, written in the momentum-helicity basis \cite{2,22}. Our results prove that generation of the Z boson and neutrinos from vacuum is possible only in the early universe when the expansion parameter $\omega$ was larger than the mass of the Z boson $M_Z$. From our computations we recover the correct Minkowski limit where the amplitude is vanishing due to the simultaneous energy and momentum conservation. The analysis in the helicity space reveals that there are nonvanishing probabilities for processes which could break the helicity conservation law and in this case the neutrino and antineutrino move along the same direction but their momenta are opposite as orientation.

The total probability was computed by considering the approximation when the Z boson momenta is small. We prove that this quantity has non-zero values only when the parameter $M_Z/\omega$ is small and in the Minkowski limit the total probability is vanishing. Further study of our analytical equations for probability can be done for obtaining the total probability in different configurations and angle fixings. The final result for the total probability shows that the divergences are of the logarithmic type, $\ln\left(\frac{M_Z}{\omega}\right)$, and that regularized quantities in de Sitter field theory will probably contain new terms that are specific to this geometry. Moreover the methods for regularization of the propagators written in closed form as a momentum integral could be completely new and we hope to approach this interesting topic in a future work.

We use here a perturbative method in which the generation of particles is the result of fields interaction in de Sitter geometry. This mechanism for matter generation receive attention only recently and seems that is suitable for studying the problem of particle production in the conditions of large expansion from early universe or strong gravitational fields. Our paper discuss a possible way of generating the gauge Z bosons and neutrinos in the early universe in processes that are forbidden in Minkowski theory by the simultaneous momentum-energy conservation. The results presented in this paper open the way for a more general study of the interactions processes with gauge bosons in Robertson-Walker space-times.

\section{Appendix}
Here we present the main steps for computing the amplitude corresponding to the generation from vacuum of triplet Z boson, neutrinos and anti-neutrino.

Using the relation that connects Hankel functions and Bessel $K$ functions \cite{21}:
\begin{equation}\label{a2}
H^{(1,2)}_{\nu}(z)=\mp \left(\frac{2i}{\pi}\right)e^{\mp
i\pi\nu/2}K_{\nu}(\mp iz),
\end{equation}
we arrive at integrals of the type \cite{21}:
\begin{eqnarray}\label{a3}
&&\int_0^{\infty} dz
z^{\mu-1}e^{-\alpha z}K_{\nu}(\beta z)=\frac{\sqrt{\pi}(2\beta)^{\nu}}{(\alpha+\beta)^{\mu+\nu}}\frac{\Gamma\left(\mu+\nu\right)\Gamma\left(\mu-\nu\right)}{\Gamma\left(\mu+\frac{1}{2}\right)}\nonumber\\
&&\times\,_{2}F_{1}\left(\mu+\nu,\nu+\frac{1}{2};\mu+\frac{1}{2};\frac{\alpha-\beta}{\alpha+\beta}\right),\nonumber\\
&&Re(\alpha+\beta)>0\,,|Re(\mu)|>|Re(\nu)|.
\end{eqnarray}
The above equation solve the temporal integrals from the amplitude.

For total probability analysis and computations, the definition for hypergeometric function was used \cite{12,18}:
\begin{eqnarray}\label{hy}
_{2}F_{1}(a,b;c;z)=1+\frac{ab}{c}\,z+\frac{a(a+1)b(b+1)}{2c(c+1)}\,z^2+\frac{a(a+1)(a+2)b(b+1)(b+2)}{6c(c+1)(c+2)}\,z^3+...
\end{eqnarray}
as well as the following identity \cite{12,18}:
\begin{eqnarray}\label{hy}
_{2}F_{1}(a,b;c;z)&=& \frac{\Gamma(c)\Gamma(c-a-b)}{\Gamma(c-a)\Gamma(c-b)}\,_{2}F_{1}(a,b;a+b-c+1;1-z)\\ \nonumber
&&+(1-z)^{c-a-b}\,\frac{\Gamma(c)\Gamma(a+b-c)}{\Gamma(a)\Gamma(b)}\,_{2}F_{1}(c-a,c-b;c-a-b+1;1-z).
\end{eqnarray}
The results for the integrals in total probability in the case $\lambda=\pm1$:
\begin{eqnarray}\label{divi}
&&\int_0^{\infty} dp\,\frac{p^3}{(\mathcal{P}+p+y)^5}=\frac{1}{4(\mathcal{P}+y)};\,\,\int_0^{\infty} dp\,\frac{p}{(\mathcal{P}+p+y)^3}=\frac{1}{2(\mathcal{P}+y)},\mathcal{P}+y>0\nonumber\\
&&\int_0^{M_Z/\omega}\frac{d\mathcal{P}\,\mathcal{P}^2}{(\mathcal{P}+y)}=\frac{1}{2}\left(\frac{M_Z}{\omega}\right)^2-y\left(\frac{M_Z}{\omega}\right)
+y^2\ln\left(\frac{M_Z}{\omega}+y\right)-y^2\ln(y);\nonumber\\
&&\int dy\left(\frac{1}{2}\left(\frac{M_Z}{\omega}\right)^2-y\left(\frac{M_Z}{\omega}\right)
+y^2\ln\left(\frac{M_Z}{\omega}+y\right)-y^2\ln(y)\right)=\frac{y}{6}\left(\frac{M_Z}{\omega}\right)^2\nonumber\\
&&-\frac{y^2}{3}\left(\frac{M_Z}{\omega}\right)+\frac{\left(\frac{M_Z}{\omega}+y\right)^3}{3}\ln\left(\frac{M_Z}{\omega}+y\right)-\frac{11}{18}\left(\frac{M_Z}{\omega}\right)^3-
y\left(\frac{M_Z}{\omega}\right)^2\ln\left(\frac{M_Z}{\omega}+y\right)\nonumber\\
&&-y^2\left(\frac{M_Z}{\omega}\right)\ln\left(\frac{M_Z}{\omega}+y\right)-\frac{y^3}{3}\ln(y).
\end{eqnarray}
The final result that give the total probability is obtained by taking the limit $y\rightarrow0$ in the last integral from equation (\ref{divi}).
In the case $\lambda=0$, the results for the integrals needed in completing the calculations for the total probability are:
\begin{eqnarray}\label{plo}
&&\int_0^{\infty} dp\,\frac{p^2}{(\mathcal{P}+p)^5}=\frac{1}{12\mathcal{P}^2}\,;\int_0^{\infty} dp\,\frac{p^2}{(\mathcal{P}+p)^6}=\frac{1}{30\mathcal{P}^3}\,;\nonumber\\
&&\int_0^{\infty} dp\,\frac{p^2}{(\mathcal{P}+p)^7}=\frac{1}{60\mathcal{P}^4}\,;\int_0^{\infty}dp\,\frac{p^2}{(\mathcal{P}+p)^4}=\frac{1}{3\mathcal{P}}.
\end{eqnarray}

The form of the helicity bispinors can be expressed as follows \cite{12,20}:
\begin{equation}\label{spin}
\xi_{\frac{1}{2}}(\vec{p}\,)=\sqrt{\frac{p_3+p}{2 p}}\left(
\begin{array}{c}
1\\
\frac{p_1+ip_2}{p_3+p}
\end{array} \right)\,,\quad
\xi_{-\frac{1}{2}}(\vec{p}\,)=\sqrt{\frac{p_3+p}{2 p}}\left(
\begin{array}{c}
\frac{-p_1+ip_2}{p_3+p}\\
1
\end{array} \right)\,,
\end{equation}
while $\eta_{\sigma}(\vec{p}\,)= i\sigma_2
[\xi_{\sigma}(\vec{p}\,)]^{*}$. These spinors satisfy the relation:
\begin{equation}\label{pa}
\vec{\sigma}\vec{p}\,\xi_{\sigma}(\vec{p}\,)=2p\sigma\xi_{\sigma}(\vec{p}\,)
\end{equation}
with $\sigma=\pm\frac{1}{2}$, where $\vec{\sigma}$ are the Pauli
matrices and $p=\mid\vec{p}\mid$ is the modulus of the momentum vector. Then the form of $\eta_{\frac{1}{2}}(\vec{p}\,')$ is given by:
\begin{eqnarray}
&&\eta_{\frac{1}{2}}(\vec{p}\,')=\sqrt{\frac{p_3'+p'}{2p'}}\left(
  \begin{array}{c}
   \frac{p_1'-ip_2'}{p_3'+p'} \\
    -1 \\
  \end{array}
\right).
\end{eqnarray}

\textbf{Acknowledgements}
This work was supported by a grant of  the Romanian Ministry of Research and Innovation, CCCDI-UEFISCDI, under project "VESS, 18PCCDI/2018",  within PNCDI III.

We would like to thank to Professor Ion Cot\u {a}escu for his observations that help us to improve the manuscript. We would also like to thank to Dr. Paul Gr\u{a}vil\u{a}, Dr. Mihaela-Andreea B\u aloi and Dr. Ciprian Sporea for their observations and for reading the manuscript.

\end{document}